\documentclass[a4paper,fleqn,usenatbib,useAMS]{mnras}

\usepackage{graphicx}	    
\usepackage{amsmath}	    
\usepackage{amssymb}	    
\usepackage{multicol}       
\usepackage{bm}		        
\usepackage{pdflscape}	    
\usepackage{multirow}       
\usepackage{lineno}         
\usepackage{xcolor}         

\newcommand{\ts}{\textsuperscript}
\newcommand{\LCDM}{$\Lambda$CDM}
\renewcommand{\Vec}[1]{\boldsymbol{#1}}
\defcitealias{fritz}{F18}
\graphicspath{{figures/}}

\newcommand{\Ref}{}
\newcommand{\Post}{}

\usepackage[T1]{fontenc}
\usepackage{ae,aecompl}

\usepackage{newtxtext,newtxmath}

\title[Anisotropy of the Milky Way satellites]{The velocity anisotropy of the Milky Way satellite system}

\author[A.H.~Riley et al.]{Alexander H.~Riley,$^{1}$\thanks{E-mail: \href{mailto:alexriley@tamu.edu}{alexriley@tamu.edu}. Code for this work is available \href{https://github.com/ahriley/beta-MW-dwarfs}{on Github}.}\href{https://orcid.org/0000-0001-5805-5766}{\includegraphics[scale=0.1]{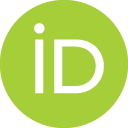}} Azadeh Fattahi,$^{2}$ Andrew B.~Pace,$^{1}$\thanks{Mitchell Astronomy Fellow} Louis E.~Strigari,$^{1}$ \newauthor Carlos S.~Frenk,$^{2}$ Facundo A.~G{\'o}mez,$^{3,4}$ Robert J.~J.~Grand,$^{5,6,7}$ \newauthor Federico Marinacci,$^{8}$ Julio F.~Navarro,$^{9}$ R{\"u}diger Pakmor,$^{7}$  \newauthor Christine M.~Simpson,$^{10,11}$ and Simon D.~M.~White$^{7}$
\\
$^{1}$Department of Physics and Astronomy, Mitchell Institute for Fundamental Physics and Astronomy, Texas A\&M University, College Station, TX 77843, USA \\
$^{2}$Institute for Computational Cosmology, Department of Physics, Durham University, South Road, Durham DH1 3LE, UK \\
$^{3}$Instituto de Investigaci{\'o}n Multidisciplinar en Ciencia y Tecnolog{\'i}a, Universidad de La Serena, Ra{\'u}l Bitr{\'a}n 1305, La Serena, Chile \\
$^{4}$Departamento de F{\'i}sica y Astronom{\'i}a, Universidad de La Serena, Av. Juan Cisternas 1200 N, La Serena, Chile \\
$^{5}$Heidelberger Institut f{\"u}r Theoretische Studien, Schlo{\ss}-Wolfsbrunnenweg 35, 69118 Heidelberg, Germany \\
$^{6}$Zentrum f{\"u}r Astronomie der Universit{\"a}t Heidelberg, Astronomisches Recheninstitut, M{\"o}nchhofstr. 12-14, 69120 Heidelberg, Germany \\
$^{7}$Max-Planck-Institut f{\"u}r Astrophysik, Karl-Schwarzschild-Str. 1, D-85748, Garching, Germany \\
$^{8}$Harvard-Smithsonian Center for Astrophysics, 60 Garden Street, Cambridge, MA 02138, USA \\
$^{9}$Department of Physics and Astronomy, University of Victoria, Victoria, BC, V8P 1A1, Canada \\
$^{10}$Enrico Fermi Institute, The University of Chicago, Chicago, IL 60637, USA \\
$^{11}$Department of Astronomy \& Astrophysics, University of Chicago, Chicago, IL 60637, USA}

\date{Accepted \Post{2019 April 4}. Received \Post{2019 March 24}; in original form \Post{2018 October 24}}

\pubyear{2019}

\begin{document}
\label{firstpage}
\pagerange{\pageref{firstpage}--\pageref{lastpage}}
\maketitle

\begin{abstract}
We analyse the orbital kinematics of the Milky Way (MW) satellite system utilizing the latest systemic proper motions for 38 satellites based on data from {\it Gaia} Data Release 2.  Combining these data with distance and line-of-sight velocity measurements from the literature, we use a likelihood method to model the velocity anisotropy, $\beta$, as a function of Galactocentric distance and compare the MW satellite system with those of simulated MW-mass haloes from the APOSTLE and Auriga simulation suites. The anisotropy profile for the MW satellite system increases from $\beta$\,$\sim$\,$-2$ at $r$\,$\sim$\,20 kpc to $\beta$\,$\sim$\,0.5 at $r$\,$\sim$\,200 kpc, indicating that satellites closer to the Galactic centre have tangentially-biased motions while those farther out have radially-biased motions. The motions of satellites around APOSTLE host galaxies are nearly isotropic at all radii, while the $\beta(r)$ profiles for satellite systems in the Auriga suite, whose host galaxies are substantially more massive in baryons than those in APOSTLE, are more consistent with that of the MW satellite system. This shape of the $\beta(r)$ profile may be attributed to the central stellar disc preferentially destroying satellites on radial orbits, or intrinsic processes from the formation of the Milky Way system. 
\end{abstract}



\section{Introduction} \label{sec:intro}

\begin{table*}
\centering
	\begin{tabular}{lcllc}
	\hline \hline
    PM study & $N_\text{sats}$ & Satellites considered & Methodology & $N_\text{stars}$ \\
    \hline
    \citet{helmi} & 12 & classical satellites, Bo{\"o}tes I & Iterative, using only DR2 data & 115/339.5/23109 \\
    \citet{simon} & 17 & $M_V > -8, d_\odot < 100$ kpc, w/ RV & Match RV members in DR2 & 2/8/68 \\
    \citet{fritz} (\citetalias{fritz}) & 39 & $r<420$ kpc w/ RV & Match RV members in DR2 & 2/18/2527 \\
    \citet{kallivayalil} & 13 & possible satellites of LMC & Iterative, initial from RV \citepalias{fritz} & 3/11/110 \\
    \citet{massari} & 7 & $M_V<-2.5$, $d_\odot < 70$ kpc & Iterative, initial from RV or HB & 29/53/189\\
    \citet{pace} & 14 & satellites in DES footprint & Probabilistic, incorporated DES DR1 & 5/15.5/67 \\
    \hline \hline
  	\end{tabular}
\caption{Summary of {\it Gaia} DR2 proper motion studies used in this analysis. $N_\text{sats}$ is the number of satellites for which a proper motion was reported in the study. $N_\text{stars}$ is the minimum/median/maximum number of stars for the list of satellites in the study. RV refers to spectroscopic radial velocity data, HB to photometric horizontal branch data. See Section \ref{sec:data} for further information.}
\label{tab:studies}
\end{table*}

\par Our presence within the Local Group offers it a special importance in astronomy.  It is the only part of the Universe where we can detect small ($M_\ast \lesssim 10^5$ M$_\odot$) dwarf galaxies, resolve their stellar populations, and study their internal properties and kinematics. As the most dark-matter dominated galaxies in the Universe \citep{2012AJ....144....4M}, these dwarfs provide crucial tests of the current structure formation paradigm -- cold dark matter with a cosmological constant (\LCDM).

\par While several predictions of \LCDM\ (e.g.~large scale structure, temperature anisotropies of the cosmic microwave background) agree with observations extraordinarily well \citep{2005Natur.435..629S, 2012AnP...524..507F}, the model faces a number of challenges on the scales of dwarf galaxy satellites (see \citealt{2017ARA&A..55..343B} for a recent review). \Ref{Many of these challenges, including the so-called missing satellites \citep{1999ApJ...522...82K, 1999ApJ...524L..19M} and too-big-to-fail \citep{2011MNRAS.415L..40B, 2012MNRAS.422.1203B} problems, have potential solutions through the inclusion of galaxy formation physics \citep{2000ApJ...539..517B, 2002MNRAS.333..156B, 2002ApJ...572L..23S} that have been reinforced by cosmological hydrodynamic simulations of galaxy formation \citep{2009MNRAS.399L.174O, 2014ApJ...786...87B, 2016MNRAS.457.1931S, 2016ApJ...827L..23W, 2018MNRAS.478..548S, 2018arXiv180604143G}.}

\par \Ref{A growing challenge to \LCDM\ is that a large fraction of satellites seem to be located on a co-rotating plane around their host galaxy (the plane-of-satellites problem; see \citealt{2018MPLA...3330004P} for a recent review). Such planes have been observed around the Milky Way \citep[MW;][]{1976MNRAS.174..695L, 2005A&A...431..517K, 2005MNRAS.363..146L, 2013MNRAS.435.2116P, fritz}, Andromeda \citep{2013Natur.493...62I}, and galaxies outside of the Local Group \citep{2014Natur.511..563I, 2015ApJ...805...67I, 2018Sci...359..534M}. The degree to which planes of satellites pose a challenge to \LCDM\ is contested; some analyses have concluded that a thin planar configuration of satellites is extremely unusual in \LCDM\ \citep{2012MNRAS.423.1109P, 2015ApJ...800...34G}, but a detailed statistical analysis (taking account of the ``look elsewhere effect'') suggests that thin satellite planes like that of the MW and M31 occur in about 10\% of galactic systems \citep{2015MNRAS.452.3838C}.}

\par Studies of the planes of satellites generally focus on two aspects of satellite kinematics: the clustering of orbital poles and the reconstruction of satellite orbits. The orbital poles of the MW satellites are more clustered than an isotropic distribution, with strong clustering measured for 8 of the 11 classical satellites \citep{2013MNRAS.435.2116P}. Orbit reconstruction is more challenging since the outcome is sensitive to the total and radial distribution of mass in the Milky Way, which are uncertain (\citealt{2015ApJS..216...29B, 2017MNRAS.465...76M}; see Figure~7 of \citealt{2019MNRAS.484.5453C} for a comparison of recent measurements of the total mass). This translates into large uncertainties in the reconstructed orbits, \Ref{making direct comparisons to theoretical predictions more complicated.}

\par To study the orbit structure of the satellite population in a potential-independent way, \citet{cautun} used the velocity anisotropy, $\beta$, to characterize the orbital properties of the satellites.  Introduced by \citet{1980MNRAS.190..873B} to quantify the orbital structure of a spherical system, $\beta$ is most commonly used in spherical Jeans equation modeling to recover the mass of a system. In a Galactocentric spherical coordinate system where $r$ corresponds to radial distance, $\theta$ the polar angle, and $\phi$ the azimuthal angle, $\beta$ is defined as:

\begin{equation}\label{eqn:beta}
    \beta(r) = 1 - \frac{\sigma_\theta(r)^2 + \sigma_\phi(r)^2}{2\sigma_r(r)^2}
\end{equation}
where $\sigma_r, \sigma_\theta, \sigma_\phi$ are the velocity dispersions along each coordinate direction. The $\beta$ parameter can take values in the range $-\infty$ to 1, where $\beta = 1$ corresponds to radial orbits plunging in and out of the Galactic centre, $\beta \rightarrow -\infty$ to circular orbits, and $\beta = 0$ to velocities being isotropically distributed at each point.

\par \Ref{Studies of $\beta$ for the MW have predominantly used either halo stars \citep{2012MNRAS.424L..44D, 2013ApJ...766...24D, 2016ApJ...820...18C, 2018arXiv181012201C} or globular clusters \citep{2018ApJ...862...52S, 2019ApJ...873..118W, 2019MNRAS.484.2832V} as tracers. While these studies have largely focused on obtaining a single value of $\beta$ as an input for spherical Jeans equation modeling, the radial anisotropy profile $\beta(r)$ also contains interesting information on the accretion history of the MW. For example, \citet{2018ApJ...853..196L} used high-resolution cosmological hydrodynamic simulations to find that dips in the $\beta(r)$ profile of halo stars may be associated with localized perturbations from, or remnants of, destroyed satellites. Such perturbations have been observed in the $\beta(r)$ profile for halo stars in the Milky Way \citep{2018arXiv181012201C}.}

\par \Ref{To date, the only measurement of $\beta$ using MW satellite galaxies has been from \citet{cautun}. Using proper motions derived using the Hubble Space Telescope (HST) for the 10 brightest satellite galaxies, they obtained $\beta = -2.2 \pm 0.4$. The low number of MW satellites with measured proper motions prohibited further studies of the $\beta(r)$ profile until the second data release from {\it Gaia} \citep[DR2;][]{2018A&A...616A...1G}. Since {\it Gaia} DR2, proper motions for nearly every MW satellite galaxy have been measured, which now motivates further studies of $\beta$ using the MW satellites as tracers.}

\par In this paper, we compare the kinematics of the Milky Way satellites to expectations from \LCDM\ using state-of-the-art cosmological hydrodynamic zoom \citep{1993ApJ...412..455K, 1996ApJ...472..460F, 2014MNRAS.437.1894O} simulations: APOSTLE \citep[A Project Of Simulating The Local Environment;][]{2016MNRAS.457..844F, 2016MNRAS.457.1931S} and Auriga \citep{2017MNRAS.467..179G}.  By focusing on $\beta$, our results only depend on the present-day kinematics of the MW satellites and not the total or radial distribution of MW mass. We use the latest satellite proper motion measurements as deduced from \Ref{{\it Gaia} DR2}, increasing the number of satellites used in an anisotropy analysis from the 10 in \citet{cautun} to 38. Furthermore, we utilize a likelihood method to determine the intrinsic $\sigma_i$'s of the MW satellite system. This more robust method, combined with the increased number of satellites spread over a wide range of Galactocentric distances ($\sim15 - 250$~kpc), allows us to perform the first measurement of $\beta(r)$ for the satellites of the Milky Way.

\par This paper is organized as follows. In Section \ref{sec:data} we review the new {\it Gaia} DR2 proper motions for MW satellites included in our analysis. In Section \ref{sec:sims} we describe the cosmological hydrodynamic zoom simulations that provide our predictions within \LCDM. In Section \ref{sec:analysis} we detail our methodology for computing $\beta$. In Section \ref{sec:results} we present the main results of our analysis and in Section \ref{sec:discussion} we provide a possible interpretation of these results. In Section \ref{sec:conclusions} we present our conclusions.

\section{Proper motions} \label{sec:data}

\par The public release of {\it Gaia} DR2 has profoundly impacted near-field cosmology in a very short period of time.  The data release contains an all-sky catalog of the five-parameter astrometric solution (position on the sky, parallax, and proper motion) for more than 1.3 billion sources \citep{2018A&A...616A...1G}. These data have already been used in multiple studies of the kinematics of the Milky Way's stellar halo \citep[e.g.][]{2018ApJ...862L...1D}, satellites \citep[e.g.][]{2019MNRAS.484.5453C} and globular clusters \citep[e.g.][]{2019MNRAS.484.2832V}.

\par We use results from six studies (see Table \ref{tab:studies} for a summary) which derive {\it Gaia} DR2 proper motions for MW satellites with comparable precision to those derived using the Hubble Space Telescope \citep[for a review of proper motions with HST, see][]{2014ASPC..480...43V}. \citet{helmi} demonstrated {\it Gaia} DR2's ability to constrain proper motions for the Magellanic Clouds, the classical (pre-SDSS) satellites, and ultra-faint dwarf Bo{\"o}tes I. \citet{simon} presented the first proper motions for many nearby ($<$\,100 kpc) ultra-faint dwarf galaxies, while \citet[][hereafter \citetalias{fritz}]{fritz} extended the limit out to 420 kpc with the largest sample of 39 satellites. \citet{kallivayalil} derived proper motions for satellites located near the Magellanic Clouds, motivated by the possibility that some of them may be satellites of the LMC itself. \citet{massari} computed proper motions for seven dwarfs, three of which do not have spectroscopic information. \citet{pace} presented a probabilistic method of determining systemic proper motions that utilized the superb photometry from the first public data release of the Dark Energy Survey \citep{2018ApJS..239...18A}.

\par The full list of satellites that we consider in this analysis is presented in Table \ref{tab:properties}, along with a summary of their properties. For this analysis we only consider satellites out to 300 kpc from the Galactic centre. We omit globular clusters and satellites whose nature is still under debate \citep[e.g.~Crater I;][]{2015ApJ...810...56K, 2016MNRAS.460.3384V}. We also do not consider overdensities that are thought to be tidally disrupting dwarf galaxies: Bo{\"o}tes III \citep{2009ApJ...702L...9C, 2018ApJ...865....7C}, Canis Major \citep{2004MNRAS.348...12M}, and Hydra I \citep{2016ApJ...818...39H}. Furthermore, we restrict our analysis to satellites which have published line-of-sight velocities in order to have full 6-D kinematic information.

\subsection{Galactocentric coordinates}\label{subsec:coords}

\begin{figure}
    \includegraphics[width=\linewidth]{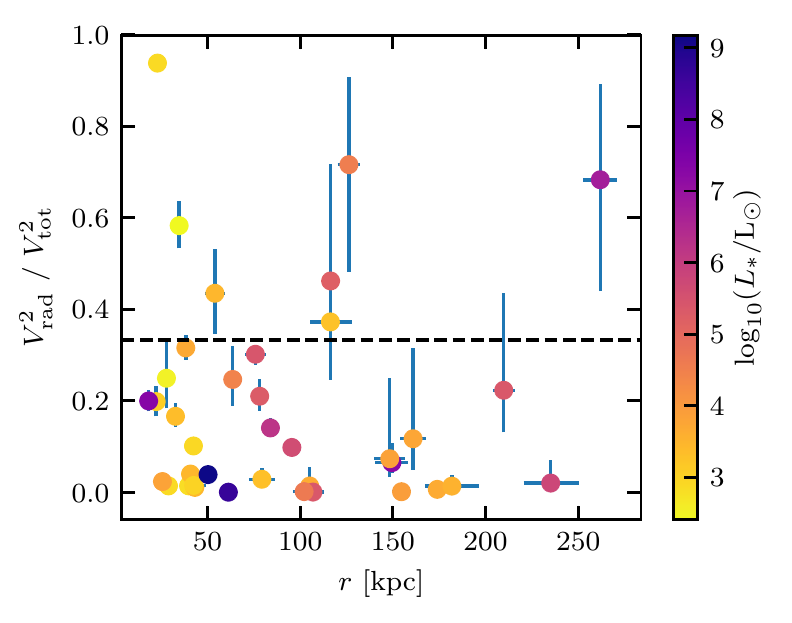}
    \caption{Tangential velocity excess of Milky Way satellites using proper motions from \citetalias{fritz} \citep[but LMC and SMC proper motions from][]{helmi}. A ratio of radial to total kinetic energy $V_\text{rad}^2/V_\text{tot}^2 \lesssim 1/3$ indicates a tangentially biased motion.} \label{fig:excess}
\end{figure}

\par In order to convert the line-of-sight velocity and proper motion measurements into Galactocentric coordinates, we use the distance measurements from Table \ref{tab:properties}. The Galactocentric Cartesian coordinates are then computed assuming a distance from the Sun to the Galactic centre of $8.2\pm0.1$ kpc, a height of the Sun relative to the Galactic plane of $25\pm5$ pc, and a solar motion relative to the Galactic centre of ($10\pm1$, $248\pm3$, $7\pm0.5$)~km~s$^{-1}$ \citep{2016ARA&A..54..529B}, in a frame where the x-axis points from the position of the Sun projected on to the Galactic plane to the Galactic centre, the y-axis points towards Galactic longitude $l=90^\circ$ (i.e.~in the direction of Galactic rotation), and the z-axis points towards the North Galactic Pole. This right-handed Cartesian system is then converted into spherical coordinates, with $r$ the distance from the Galactic centre, polar angle $\theta$ defined from the z-axis, and azimuthal angle $\phi$ defined from the x-axis such that the Galactic rotation is in the $-\phi$ direction.

\par We perform 2,000 Monte Carlo simulations drawing satellite proper motions, line-of-sight velocities, and heliocentric distances randomly from Gaussian distributions centred on their measured values with dispersions given by their respective errors.  When drawing the proper motions we account for the correlation between $\mu_{\alpha*} \equiv \mu_\alpha\cos\delta$ and $\mu_\delta$ if provided in the proper motion study. The randomly drawn kinematic properties are then converted into Galactocentric spherical coordinates as described in the previous paragraph. The resulting Galactocentric positions and velocities (and corresponding uncertainties), obtained directly from the observed distance, line-of-sight velocity, and proper motion measurements, are summarized in Table \ref{tab:galactoprops}. 

\par \Ref{To illustrate the tangential nature of the motions of the MW satellites, we show the ratio of of radial to total kinetic energy $V_\text{rad}^2/V_\text{tot}^2$ for each satellite in Figure \ref{fig:excess}. A ratio $\lesssim$ 1/3 indicates a tangentially-biased motion. We find that $\sim$80\% of MW satellites show a tangential velocity excess, comparable to \citet{cautun} who found that 9 of the 10 brightest MW satellites had tangentially-biased motions.}

\subsection{Sample selections}\label{subsec:PMstudy}

\par It is important to note that the proper motions derived by both \citet{simon} and \citetalias{fritz} were based only on matching spectroscopically confirmed member stars with {\it Gaia} DR2 data, in some cases depending on very few ($N\sim2-5$) stars to derive a systemic proper motion. The small number statistics could lead to a biased result; \citet{massari} found that the subsample of spectroscopic members in Segue 2 used by \citet{simon} and \citetalias{fritz} is systematically shifted in proper motion space relative to the full sample recovered using their iterative method. To avoid problems from this potential bias, as well as to confirm that our results do not depend strongly on the systematics associated with a particular study, we consider three different samples of proper motion data in our analysis:
\begin{enumerate}
    \item \Ref{38 satellites, comprised of 36 satellites from \citetalias{fritz} plus the LMC and SMC. This is the full list of satellites in Table \ref{tab:properties}.}
    \item a ``gold'' sample constructed by prioritizing studies which included steps in their analysis to increase the sample of member stars beyond the spectroscopic sample. For example, \citet{pace} used a probabilistic method incorporating photometry from the first public data release of the Dark Energy Survey \citep{2018ApJS..239...18A}. This ``gold'' sample consists of 32 satellites,\footnote{\Ref{The satellites that are in the full sample that are excluded from the ``gold'' sample are: Canis Venatici I, Canis Venatici II, Hercules, Leo IV, Leo V, and Pisces II.}} with proper motions taken from the five other previous studies.
    \item the same 32 satellites from the ``gold'' sample, but using the \citetalias{fritz} proper motions.
\end{enumerate}

\par Since {\it Gaia} DR2 proper motions for the Magellanic Clouds have only been reported by \citet{helmi}, we use these proper motion measurements in all samples. The exact study used for each satellite in the ``gold'' sample is shown in Table \ref{tab:properties}.  As detailed in Section \ref{sec:results}, we find that our results do not depend on which sample is used. We focus on the results for the full 38 satellite sample using \citetalias{fritz} proper motions in the following sections.

\section{Cosmological simulations} \label{sec:sims}

\begin{table*}
\centering
	\begin{tabular}{cccccccc}
	\hline \hline
    Run & $M_{200}$ & $R_{200}$ & $M_\ast$ & $R_{1/2,\ast}$ & $V_\text{circ}(8.2\text{ kpc})$ & $N_\text{subs}$ & $N_\text{subs}$ \\
    & [$10^{12}$ M$_\odot$] & [kpc] & [$10^{10}$ M$_\odot$] & [kpc] & [km s$^{-1}$] & ($V_\text{max} > 5$ km s$^{-1}$) & ($M_\ast > 0$) \\
    \hline
    \multicolumn{8}{c}{APOSTLE} \\
    \hline
    \multirow{2}{*}{AP-01} & 1.57 & 238.8 & 2.3 & 8.2 & 173.6 & 1187 & 52 \\
    & 1.02 & 206.8 & 1.1 & 6.2 & 124.3 & 1071 & 41 \\
    \multirow{2}{*}{AP-04} & 1.16 & 216.2 & 1.1 & 5.0 & 155.3 & 1006 & 63 \\
    & 1.12 & 213.7 & 1.6 & 4.7 & 148.8 & 1232 & 62 \\
    \multirow{2}{*}{AP-06} & 2.04 & 260.6 & 2.2 & 5.5 & 172.6 & 1517 & 76 \\
    & 1.07 & 210.3 & 1.2 & 7.4 & 136.7 & 999 & 27 \\
    \multirow{2}{*}{AP-10} & 1.43 & 231.5 & 2.2 & 6.7 & 163.6 & 1105 & 35 \\
    & 0.47 & 160.1 & 1.0 & 6.2 & 121.0 & 669 & 26 \\
    \multirow{2}{*}{AP-11} & 0.90 & 198.5 & 1.0 & 3.3 & 150.4 & 810 & 36 \\
    & 0.78 & 189.3 & 0.9 & 4.2 & 136.5 & 784 & 33 \\
    \hline
    \multicolumn{8}{c}{Auriga} \\
    \hline
    Au6 & 1.01 & 211.8 & 6.3 & 4.7 & 224.2 & 517 & 74 \\
    Au16 & 1.50 & 241.5 & 8.8 & 9.6 & 217.7 & 594 & 95 \\
    Au21 & 1.42 & 236.7 & 8.6 & 7.7 & 231.7 & 621 & 97 \\
    Au23 & 1.50 & 241.5 & 8.8 & 8.1 & 240.6 & 582 & 83 \\
    Au24 & 1.47 & 239.6 & 8.5 & 8.4 & 219.0 & 629 & 87 \\
    Au27 & 1.70 & 251.4 & 9.7 & 6.6 & 254.5 & 564 & 104 \\
    \hline
    MW & $1.1\pm0.3$ & 220.7$^a$ & $5\pm1$ & -- & $238\pm15$ & -- & $124^{+40}_{-27}$ \\
    \hline \hline
  	\end{tabular}
\caption{Summary of simulation parameters at $z=0$. The columns are: 1) simulation name, 2) \Ref{halo} mass, 3) \Ref{halo} radius, 4) stellar mass within $0.15\times R_{200}$, 5) half-stellar mass radius, 6) circular velocity at $R_0=8.2$ kpc, 7) number of subhaloes with maxiumum circular velocity $V_\text{max} > 5$ km s$^{-1}$, and 8) number of subhaloes which form stars. Note that the APOSTLE volumes contain a pair of haloes as analogs for the MW and M31. Each halo in a pair is considered separately in this analysis. The final row provides current estimates for these quantities for the Milky Way from \citet{2016ARA&A..54..529B} and \citet{2018MNRAS.479.2853N}, though note that the latter considers satellites down to $M_V = 0$, too faint for APOSTLE and Auriga to resolve. $^a$ refers to the mean of the values for $R_{200}$ provided in Table 8 of \citet{2016ARA&A..54..529B}, the standard deviation of which is 28.6 kpc. The value for the mass of the MW is consistent with the latest measurement from {\it Gaia} DR2 halo globular cluster motions \citep[$1.41_{-0.52}^{+1.99}\times10^{12}$~M$_\odot$;][]{2019ApJ...873..118W}.}
\label{tab:simprops}
\end{table*}

\par To compare our results with the expectations from the standard \LCDM\ cosmology, we utilize suites of self-consistent cosmological hydrodynamic zoom simulations of Local Group analogs, APOSTLE \citep{2016MNRAS.457..844F, 2016MNRAS.457.1931S}, and of Milky Way analogs, Auriga \citep{2017MNRAS.467..179G}. These two simulation suites have similar resolution and include similar baryonic processes (e.g.~star formation, stellar, supernova and black hole feedback, uniform background UV field for reionization), though the numerical methods and prescriptions for subgrid physics are different (see references in following subsections for details). We also analyse dark-matter-only runs from these suites for comparison.

\subsection{APOSTLE}

\par The APOSTLE project is a suite of cosmological hydrodynamic zoom simulations of twelve volumes using the code developed for the EAGLE project \citep{2015MNRAS.446..521S, 2015MNRAS.450.1937C}. The galaxy formation model includes metallicity-dependent star formation and cooling, metal enrichment, stellar and supernova feedback, homogeneous X-ray/UV background radiation (hydrogen reionization assumed at $z_{reion} = 11.5$), supermassive black-hole formation and growth, and AGN activity \citep{2009MNRAS.398...53B, 2015MNRAS.454.1038R}. The full details of the subgrid physics can be found in \citet{2015MNRAS.446..521S}.

\par The APOSTLE volumes were selected to have similar dynamical properties as the Local Group; the full selection procedure is described in \citet{2016MNRAS.457..844F} and a discussion of the main simulation characteristics is given in \citet{2016MNRAS.457.1931S}. In summary, each volume consists of a MW/M31-like pair of haloes with \Ref{halo} mass\footnote{Defined to be the mass inside a sphere in which the mean matter density is 200 times the critical density $\rho_\text{crit} = 3H^2(z)/8\pi G$. \Ref{Virial quantities are defined at that radius and are identified by a `200' subscript.}} ranging from $0.5-2.5\times10^{12}$ M$_\odot$, separated by $800\pm200$ kpc, approaching with radial velocity $<250$~km~s$^{-1}$ and tangential velocity $< 100$~km~s$^{-1}$.  The haloes are isolated, with no additional halo larger than the smaller of the pair within 2.5 Mpc of the midpoint between the pair, and in environments where the Hubble flow is relatively unperturbed out to 4 Mpc. \Ref{The simulations adopt the WMAP-7 cosmological parameters \citep{2011ApJS..192...18K}: $\Omega_M = 0.272$, $\Omega_\Lambda = 0.728$, $h = 0.704$, $\sigma_8 = 0.81$, and $n_s = 0.967$.}

\par The volumes were simulated at three resolution levels, the highest of which (and the only level considered here) has primordial gas (DM) particle masses approximately $1.0(5.0)\times10^4$ M$_\odot$, with a maximum force softening length of 134 pc. Five volumes have been simulated so far at this resolution, corresponding to AP-01, AP-04, AP-06, AP-10, and AP-11 in Table 2 of \citet{2016MNRAS.457..844F}. Each halo in a pair is treated separately in this analysis, resulting in ten high-resolution APOSTLE haloes being considered in this work.

\subsection{Auriga}

\par The Auriga simulations \citep{2017MNRAS.467..179G} are a suite of cosmological magnetohydrodynamic zoom simulations of single MW-like haloes with \Ref{halo} masses in the range $1-2\times10^{12}$ M$_\odot$. They were performed with the moving mesh code AREPO \citep{2010MNRAS.401..791S} and a galaxy formation model that includes primordial and metal line cooling, a prescription for a uniform background UV field for reionization (completed at $z = 6$), a subgrid model for star formation and stellar feedback \citep{2003MNRAS.339..289S}, magnetic fields \citep{2014ApJ...783L..20P, 2017MNRAS.469.3185P}, and black hole seeding, accretion, and feedback.

\par The Auriga haloes were identified as isolated in the $z=0$ snapshot of the parent dark-matter-only simulation with a 100 Mpc box-side length of the EAGLE project introduced in \citet{2015MNRAS.446..521S}. \Ref{To be considered isolated, the centre of any target halo must be located outside 9 times the \Ref{halo} radius of any other halo that has a mass greater than 3\% that of the target.} The simulations assumed the \citet{2014A&A...571A..16P} cosmological parameters: $\Omega_M = 0.307$, $\Omega_\Lambda = 0.693$, $h = 0.6777$, $\sigma_8 = 0.8288$, and $n_s =  0.9611$.

\par The volumes were simulated at three resolution levels, the highest of which (and the only level considered here) has baryonic element (DM particle) masses approximately $0.5(4.0)\times10^4$M$_\odot$, with a maximum force softening length of 185 pc, comparable to the highest resolution for APOSTLE.  Thus far, six haloes have been resimulated at this high resolution, corresponding to Au6, Au16, Au21, Au23, Au24, and Au27 in Table 1 of \citet{2017MNRAS.467..179G}.

\subsection{Stellar discs}

\begin{figure}
    \includegraphics[width=\linewidth]{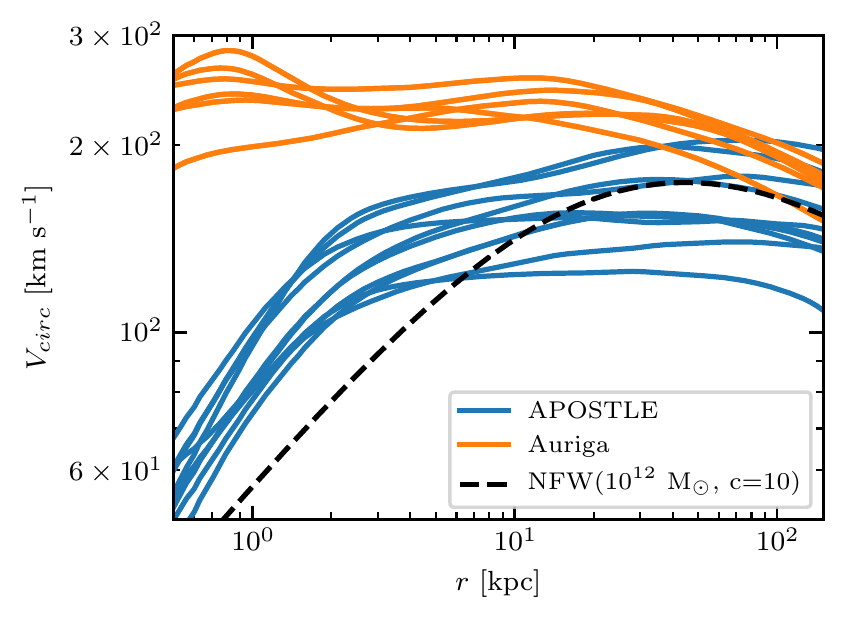}
    \caption{Circular velocity profiles based on total mass for APOSTLE (blue) and Auriga (orange) haloes. The formation of stellar discs in Auriga is reflected in a much deeper potential near the centre of the halo. For comparison, we also show the circular velocity profile for an NFW \citep{1996ApJ...462..563N, 1997ApJ...490..493N} halo with $M_{200} = 10^{12}$ M$_\odot$ and $c=10$ (black, dashed), where $c$ is the ratio between the virial radius and the NFW scale radius.} \label{fig:vcirc}
\end{figure}

\par Even though the APOSTLE and Auriga haloes have similar \Ref{halo} masses, the difference in their baryon content at $z=0$ affects the shape and depth of their potentials (and hence the dynamics of their satellite systems). A main difference between the two simulation suites is the mass and morphology of the stellar discs of their main galaxies.\footnote{The gas content of these galaxies is sub-dominant compared to the stellar component at small radii.} The Auriga simulations are able to produce radially extended and thin discs, with sizes comparable to that of the MW \citep{2017MNRAS.467..179G, 2018MNRAS.481.1726G}, while their total stellar masses are slightly higher, close to $10^{11}$~M$_\odot$, than that of the MW. By contrast, the APOSTLE host galaxies have morphologies that are less disky with relatively low stellar mass, $\sim 10^{10}$~M$_\odot$.\footnote{We note that the host galaxies in low and medium resolution APOSTLE runs have disky morphologies and higher stellar masses compared to the high resolution runs used here.}  A summary of properties for each simulation run is shown in Table \ref{tab:simprops}.
 
\par The total circular velocity profiles, $V_\text{circ}=\sqrt{G M(<r)/r}$, for the two simulations are shown in Figure \ref{fig:vcirc}. The different behaviour of the APOSTLE \Ref{and Auriga} haloes (blue \Ref{and yellow} curves) compared to the NFW circular velocity profile (black dashed curve) is due to the contraction of haloes in response to the presence of baryons. The much larger difference in the circular velocity profiles between the two simulation suites is due to the more massive stellar discs in Auriga (orange curves) combined with the enhanced dark matter contraction.

\par These differences are useful in quantifying the effect of a stellar disc on $\beta$. The deepening of the potential due to the large baryonic disc, combined with the non-spherical potential of the disc, can affect the tidal stripping of subhaloes. Hydrodynamic simulations suggest that tidal effects from \Ref{a} baryonic disc near the centre of a host halo can reduce the number of dark substructures by up to a factor of two \citep{2014ApJ...786...87B, 2016MNRAS.458.1559Z, 2017MNRAS.467.4383S, 2018ApJ...859..129N}, an effect that is reproduced in DMO simulations with an embedded disc potential \citep{2010ApJ...709.1138D, 2015MNRAS.452.2367Y, 2017MNRAS.471.1709G, 2018arXiv181112413K}. This tidal disruption preferentially affects radial orbits that come close to the disc \citep{2017MNRAS.471.1709G}, implying that surviving subhaloes in the inner regions should be on \Ref{circular} orbits, resulting in a lower $\beta$ near the centre.

\subsection{Matching the \Ref{radial distributions}} \label{subsec:matchdata}

\begin{figure}
    \includegraphics[width=\linewidth]{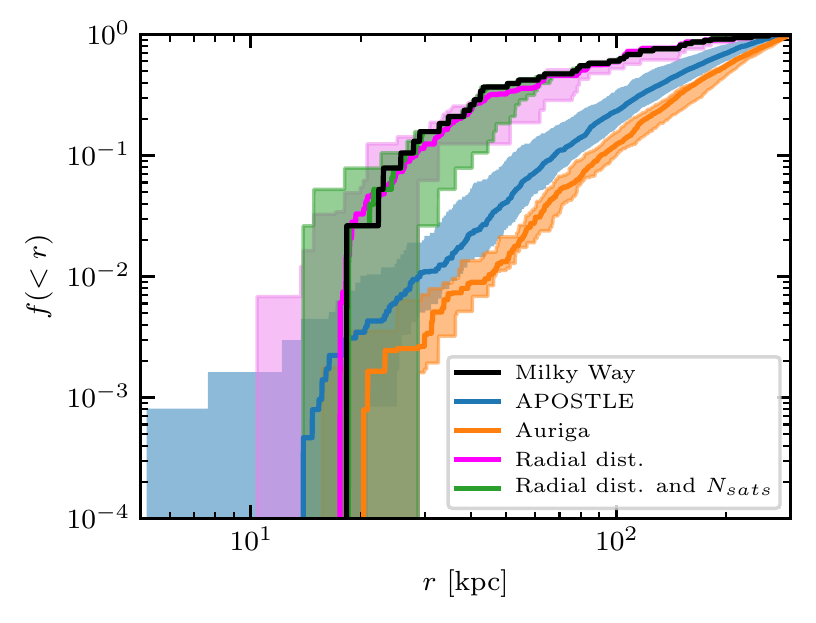}
    \caption{The radial distribution of subhaloes with $V_\text{max} > 5$ km s$^{-1}$ in APOSTLE (blue) and Auriga (orange) compared to that of the MW satellites (black). The deepening of the potential in Auriga haloes results in a less centrally concentrated radial distribution compared to APOSTLE, due to more subhaloes being destroyed. The \Ref{magenta} contours correspond to matching the radial distribution but not number of satellites\Ref{, while the green contours correspond to additionally matching the number of satellites} (see Section~\ref{subsec:matchdata} for details) for both suites. Solid curves indicate the median and shaded regions the total spread.}
    \label{fig:radialdist}
\end{figure}

\par \Ref{In addition to comparing the results of simulated systems to that of the MW satellites, we would like to select samples of subhaloes that are more representative of the observed MW satellites' radial distribution. By comparing the results of these subhalo samples to those of the full simulated systems, we will be able to understand the impact of the tracers' radial distribution on our results. We will also provide a fairer comparison between the simulated systems and the MW satellites.} 

\par We begin by considering all subhaloes which, at $z$\,$=$\,0, have maximum circular velocity $V_\text{max} > 5$~km~s$^{-1}$. \Ref{This is a conservative resolution limit for both APOSTLE and Auriga, roughly corresponding to subhalo masses of $\sim5\times10^6$ M$_\odot$ or containing $\sim100$ DM particles.} These subhaloes are a mix of dark and luminous \Ref{(i.e.~contain stars)}; typically $\sim4$\% contain stars in APOSTLE and $\sim15$\% contain stars in Auriga (see Table \ref{tab:simprops}). 

\par \Ref{We then create two subhalo samples resulting from} (1) matching the radial distribution \Ref{of subhaloes to that of the MW satellites} and (2) additionally matching the abundance of \Ref{subhaloes to that of the MW satellites}. When matching both \Ref{radial distribution and abundance (case 2)}, we simply \Ref{compare the Galactocentric distance of each MW satellite to the host-centric distance of each subhalo and select the closest match} (without replacement).  When only matching the radial distribution and not the abundance \Ref{(case 1)}, we select subhaloes based on the following inverse transform sampling method:
\begin{enumerate}
    \item Compute the cumulative distribution function (CDF) of \Ref{MW satellite Galactocentric distances}.
    \item Generate a random number uniformly between 0 and 1 and map that number to a \Ref{distance} using the CDF from step 1.
    \item Select the subhalo \Ref{that has a host-centric distance that is closest to the randomly generated distance} and add it to the sample if it is within 5 kpc of the randomly generated value.  This subhalo is removed from possible selection in the future, i.e. without replacement.
    \item Repeat steps (ii)-(iii) until a \Ref{distance} is generated that does not have a subhalo match within 5 kpc.
\end{enumerate}

\par The 5~kpc cutoff is meant to strike a balance between increasing the number of subhaloes in the sample (higher cutoff) and providing a close match \Ref{between the radial distribution of subhaloes} to \Ref{that of the} MW satellites (lower cutoff).  Our results are not sensitive to the exact value of this cutoff. Using this method, we typically find \Ref{subhalo} radial distributions that are much closer to the MW satellite distribution than that of the original subhalo populations (see Figure~\ref{fig:radialdist}).

\section{Likelihood Analysis} \label{sec:analysis}

\par We seek to model the orbital kinematics of Milky Way satellites and compare the results with those of cosmological simulations using the velocity anisotropy parameter $\beta$. Two models are considered: (1) a constant value of $\beta$ at all radii and (2) one in which $\beta$ varies as a function of Galactocentric distance.  To determine the posterior probability densities for each model, we use \texttt{emcee} \citep{emcee}, an implementation of the affine-invariant ensemble sampler for Markov chain Monte Carlo (MCMC).

\subsection{Framework and \Ref{constant $\beta$ model}} \label{subsec:uniform}

\par We assume that the velocity distribution of the MW satellite system in Galactocentric spherical coordinates ($r, \theta, \phi$) is a multivariate Gaussian with different means and dispersions in each direction.  The resulting likelihood $F_i$ for a given satellite $i$ with velocity $\Vec{v}_i = (v_{r,i}, v_{\theta, i}, v_{\phi,i})$ is then

\begin{equation}
    F_i = \frac{1}{\sqrt{(2\pi)^3 \left|\Vec{\Sigma}_i\right|}}\exp\left[-\frac{(\Vec{v}_i-\Vec{v}_\text{sys})^\text{T} \Vec{\Sigma}_i^{-1}(\Vec{v}_i-\Vec{v}_\text{sys})}{2} \right],
    \label{eqn:gaussian}
\end{equation} 
where $\Vec{v}_\text{sys} = (v_r, v_\theta, v_\phi)$ are the intrinsic mean velocities of the system \Ref{(i.e.~the entire population of MW satellites)} and the covariance matrix is

\begin{equation}
    \Vec{\Sigma}_i \equiv \begin{bmatrix}
        \sigma_r^2 + \delta_r^2 & C_{r\theta}\delta_r\delta_\theta & C_{r\phi}\delta_r\delta_\phi \\
        C_{r\theta}\delta_r\delta_\theta & \sigma_\theta^2 + \delta_\theta^2 & C_{\theta\phi}\delta_\theta\delta_\phi \\
        C_{r\phi}\delta_r\delta_\phi & C_{\theta\phi}\delta_\theta\delta_\phi & \sigma_\phi^2 + \delta_\phi^2
    \end{bmatrix}.
\end{equation}
Here, $(\sigma_r, \sigma_\theta, \sigma_\phi)$ are the intrinsic dispersions of the system \Ref{(i.e.~the entire population of MW satellites)} and $(\delta_r, \delta_\theta, \delta_\phi, C_{r\theta}, C_{r\phi}, C_{\theta\phi})$ are the observed measurement errors and correlation coefficients for the velocities of the given satellite, which are obtained from the 2,000 Monte Carlo samples described in Section \ref{subsec:coords}. 

\par Due to the conversion from heliocentric to Galactocentric spherical coordinates, the resulting satellite velocity errors are not necessarily Gaussian in each component. We find that approximating the errors as Gaussian is reasonable in most cases, though Draco II, Tucana III, and Willman 1 show significant skewness and kurtosis in both $v_\theta$ and $v_\phi$. 

\par The combined log-likelihood for the full satellite sample is then

\begin{align}
    \ln \mathcal{L} &= \sum_{i=1}^{N_\text{sats}} \ln F_i  = -\frac{1}{2}\sum_{i=1}^{N_\text{sats}} 3\ln{2\pi} + \ln{\left|{\Vec{\Sigma}_i}\right|} + \Vec{u}_i^\text{T} \Vec{\Sigma}_i^{-1}\Vec{u}_i \\
    &\propto -\sum_{i=1}^{N_\text{sats}} \Vec{u}_i^\text{T} \Vec{\Sigma}_i^{-1}\Vec{u}_i + \ln \left|\Vec{\Sigma}_i\right|, \label{eqn:likelihood}
\end{align}
where $\Vec{u}_i \equiv \Vec{v}_i - \Vec{v}_\text{sys}$. Equation \ref{eqn:likelihood} is the likelihood function used to probe the model parameter space with \texttt{emcee}.

\par The first model we consider assumes constant velocity dispersions at all radii, resulting in a constant value for $\beta$. We impose spherical symmetry by requiring $v_r = v_\theta = 0$ and $\sigma_\theta^2 = \sigma_\phi^2$\Ref{, as is commonly assumed in Jeans equation modeling of the dynamics of a system}. In total this model then has 3 free parameters: a mean rotational motion $v_\phi$ and dispersions $\sigma_r$ and $\sigma_\theta$. We assume uniform priors for the mean motion $-500 < v_\phi < 500$~km~s$^{-1}$ as well as for the dispersions $0 < \sigma_i < 300$~km~s$^{-1}$. We repeat the analysis with Jeffreys prior $-3 < \log_{10}\sigma_i < 3$ and find that this does not meaningfully change our results.

\subsection{Variable dispersions with radius} \label{subsec:variable}

\par To take advantage of the increased number of satellites with proper motions over a wide range of Galactocentric distances, we include a separate likelihood analysis in which we adopt a simple model for the velocity dispersion to vary as a function of radius in each coordinate $j$:

\begin{equation} \label{eqn:variable}
    \sigma_j(r) = {\sigma_{j,0}}{\left(1 + \frac{r}{r_{j,0}}\right)^{-\alpha_j}}
\end{equation} 
where $\sigma_{j,0}$ and $r_{j,0}$ are the characteristic dispersion and length scales and $\alpha_j$ is the slope of the fall off at large radii. We then use the same likelihood function as in Section \ref{subsec:uniform} (specifically Equation~\ref{eqn:likelihood}) with the additional parameters introduced in Equation~\ref{eqn:variable}. The $\beta(r)$ profile then follows from Equation~\ref{eqn:beta}.

\begin{figure}
    \centering
    \includegraphics[width=\linewidth]{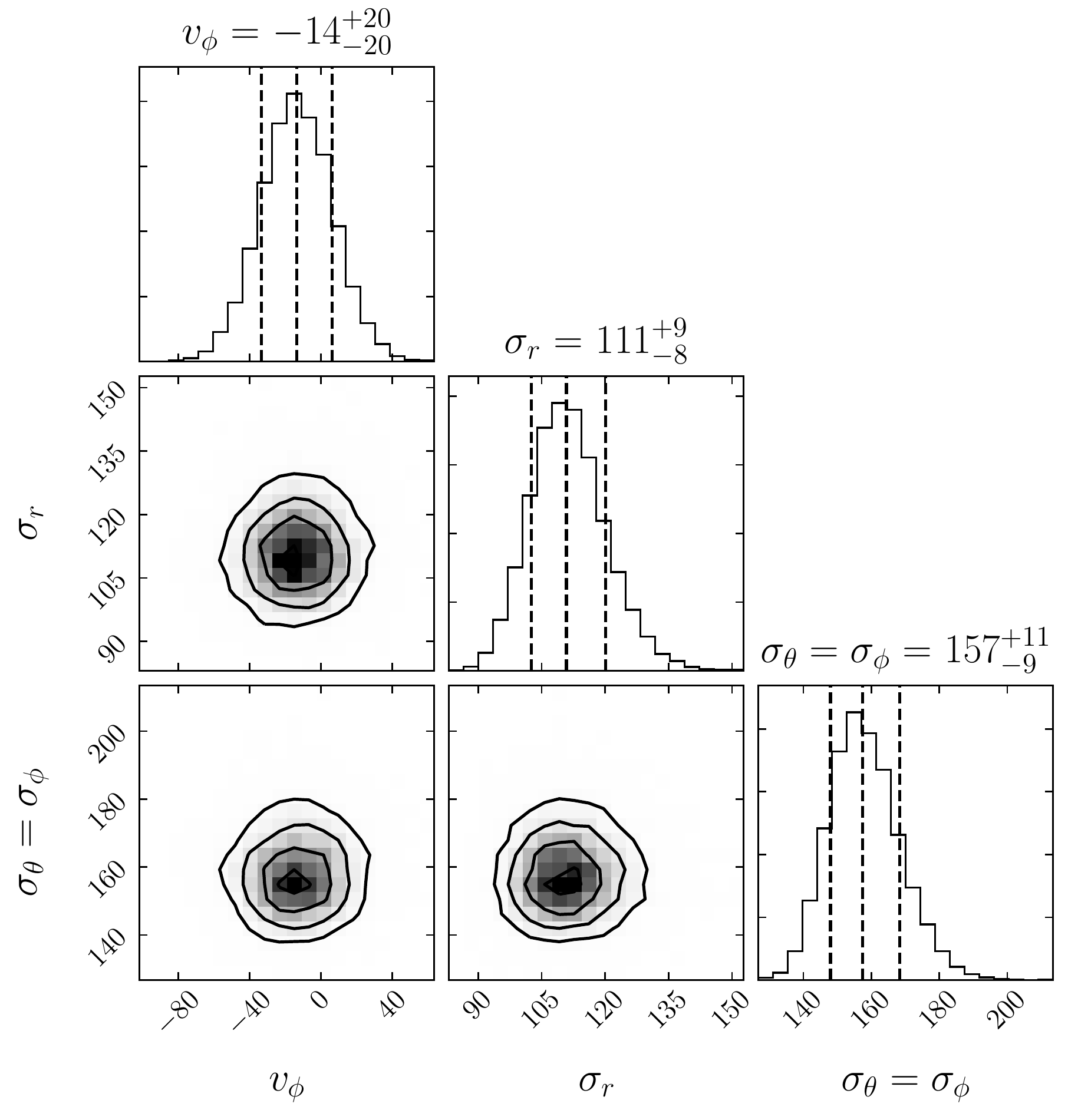}
    \caption{Posterior distributions for \Ref{all 38} Milky Way satellites assuming the \Ref{constant $\beta$ model, using \citetalias{fritz} proper motions}. From left to right the parameters are: $v_\phi$ (systemic rotational motion in km~s$^{-1}$), $\sigma_r$, and $\sigma_\theta=\sigma_\phi$ (intrinsic velocity dispersions in km~s$^{-1}$). The contours enclose 39.4, 85.5, 98.9\% of the posterior distribution corresponding to $1, 2, 3 - \sigma$ confidence intervals. The dotted lines on the 1-D histograms are the 16\ts{th}, 50\ts{th}, and 84\ts{th} percentiles and the numerical values are quoted above.}
    \label{fig:uniformcorner}
\end{figure}

\par As in the \Ref{constant $\beta$ model}, we impose spherical symmetry by requiring $v_r = v_\theta = 0$ and $\sigma_\theta^2(r) = \sigma_\phi^2(r)$. In total this model then has 7 parameters: a mean rotational motion $v_\phi$ (which is held constant with $r$), the characteristic dispersion and length scales $\sigma_{i,0}$ and $r_{i,0}$, and the slope $\alpha_i$, for both Galactocentric spherical coordinates $r$ and $\theta$. We assume the same uniform prior for the mean motion $-500 < v_\phi < 500$~km~s$^{-1}$ as in the \Ref{constant $\beta$ analysis}.  For the $\sigma_i(r)$ parameters we assume uniform priors $50 < \sigma_{i,0} < 1000$~km~s$^{-1}$, $10 < r_{i,0} < 1000$~kpc, and $0 < \alpha_i < 10$. We repeat the analysis with Jeffreys priors $-3 < \log_{10}\sigma_{i,0} < 3$ and $-3 < \log_{10} r_{i,0} < 3$ and again find that this change of priors does not meaningfully change our results.

\section{Results} \label{sec:results}

\par We now present the resulting posterior probability densities for $\beta(r)$ for the MW satellite system using the models described above. We show that satellites within $r \lesssim 100$ kpc have more tangentially-biased motions (lower $\beta$) than those farther away. This result is also seen in simulated MW analogs, but it is difficult to disentangle effects due to the central stellar disc from those imprinted at formation. From here onwards, we refer to dark-matter-only simulations from the APOSTLE and Auriga suites collectively as ``DMO" and the haloes simulated with baryonic physics by their suite name.

\subsection{\Ref{Constant $\beta$ model}}

\par
\begin{figure}
    \includegraphics[width=\linewidth]{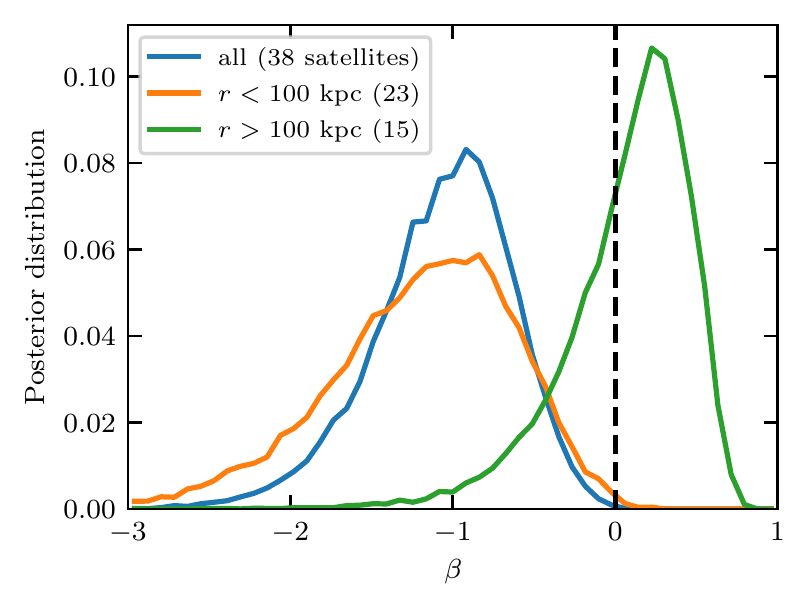}
    \caption{Posterior distributions for $\beta$ assuming the \Ref{constant $\beta$ model with \citetalias{fritz} proper motions.} The results are shown using \Ref{all 38} satellites (blue), satellites within 100 kpc (orange, \Ref{23 satellites}), and satellites outside of 100 kpc (green, \Ref{15 satellites}). The vertical black dashed line corresponds to the isotropic case $\beta=0$.}
    \label{fig:uniform}
\end{figure}

 The posterior distribution of parameters for the \Ref{constant $\beta$ model} for the MW satellites is shown in Figure~\ref{fig:uniformcorner} and the resulting posterior for $\beta$ is shown in Figure~\ref{fig:uniform} (blue curve). We find that the satellites are overall on near tangential orbits, with $\beta = -1.02_{-0.45}^{+0.37}$. We do not find significant evidence for \Ref{the MW satellite population exhibiting} rotation parallel to the plane of the Milky Way \Ref{disc} ($v_\phi = -14_{-20}^{+20}$~km~s$^{-1}$\Ref{; note that a star located in the disc would have $v_\phi$ on the order of $\sim$100 km s$^{-1}$}). \Ref{We also find that the constant $\beta$ results are similar when using the different samples described in Section \ref{subsec:PMstudy} (``gold'' sample and ``gold'' satellites with \citetalias{fritz} proper motions).}

\par To better understand the results from the variable $\beta(r)$ model, we examine two radial bins. We split the satellites into two populations, one with $r < 100$~kpc (23 satellites) and the other with $r > 100$~kpc (15 satellites), and perform the same \Ref{constant} $\beta$ analysis on each. The inner and outer regions clearly have different posterior distributions (Figure~\ref{fig:uniform}, orange and green curves respectively), with the inner region having a more negative (i.e., more tangentially biased) $\beta$ posterior than the outer region. These results do not change when considering each of the different proper motion samples described in Section \ref{subsec:PMstudy}. This supports the finding in the $\beta(r)$ model (discussed below) that satellites in the inner region ($r \lesssim 100$ kpc) have more tangentially-biased motions than those farther away.

\subsubsection{\Ref{Comparison to Cautun \& Frenk \Post{(2017)}}}

\par \Ref{Our \Ref{constant} $\beta$ result agrees within 2$\sigma$ with the result of \citet{cautun}. These authors found $\beta = -2.2 \pm 0.4$ using HST proper motions of only 10 of the brightest satellites and simply computing $\beta$ from Monte Carlo realizations of the MW satellite system using observational errors.  Using updated {\it Gaia} DR2 proper motions and our likelihood method, our result for that same subsample of 10 satellites is $\beta = -1.52_{-1.23}^{+0.86}$, which is consistent with \citet{cautun}. The small offset is likely due to different input data and analysis techniques.}

\subsection{Variable $\beta$ model}

\begin{figure}
    \includegraphics[width=\linewidth]{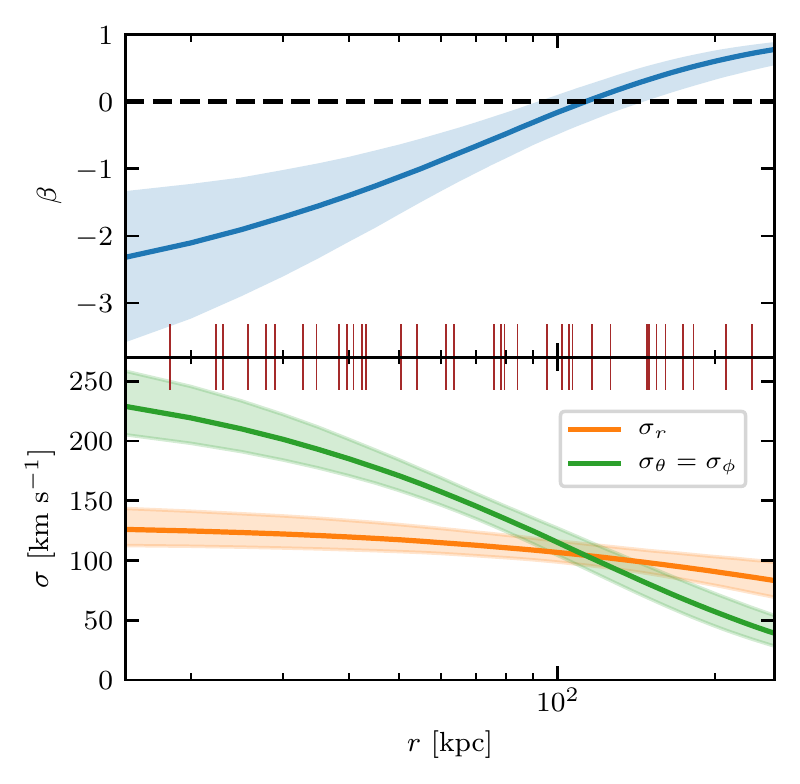}
    \caption{Posterior distributions for the $\beta(r)$ model for \Ref{the 38} MW satellites \Ref{using \citetalias{fritz} proper motions}. Top: posterior for $\beta(r)$. The horizontal black dashed line corresponds to the isotropic case, $\beta=0$. Bottom: posterior for the systemic dispersion in $\sigma_r$ (\Ref{orange}) and $\sigma_\theta = \sigma_\phi$ (\Ref{green}). The \Ref{brown} ticks along the middle axis mark the radial positions of the MW satellites. For both panels, solid curves correspond to the median values and the shaded region the $16-84$\% confidence interval.}
    \label{fig:variable}
\end{figure}

\par The posterior distribution for the parametrized $\beta(r)$ model \Ref{for the MW satellites} is shown in Figure~\ref{fig:variable} \Ref{(top panel), along with the dispersions $\sigma_r$ and $\sigma_\theta = \sigma_\phi$ (bottom panel)}. We find that the radial profile dips in the inner ($<100$ kpc) region to $\beta$\,$\sim$\,$-2$ at $r\sim20$ kpc and flattens out to $\beta$\,$\sim$\,0.5 in the outer region. This again indicates that satellites near the centre of the Milky Way have tangentially-biased motions, while satellites in the outer region have more radially-biased motions.

\begin{figure}
    \includegraphics[width=\linewidth]{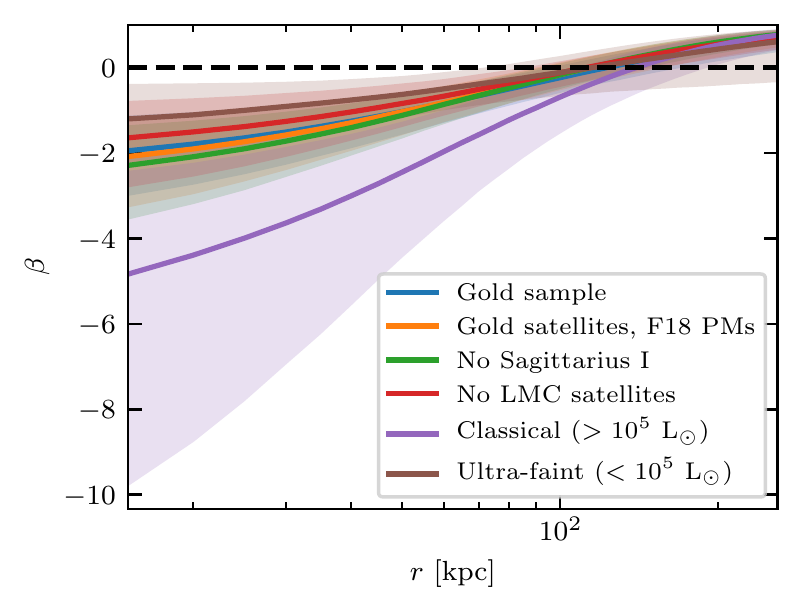}
    \caption{\Ref{Posterior distributions for the $\beta(r)$ model using different satellite samples. The ``gold'' sample and ``gold'' satellites using \citetalias{fritz} proper motions are as described in Section \ref{subsec:PMstudy}. ``No LMC satellites'' excludes the candidate satellites of the LMC as identified by \citet{kallivayalil}. We define ultra-faint dwarf galaxies as those fainter than $10^5$ L$_\odot$ \citep{2017ARA&A..55..343B} and refer to galaxies brighter than this limit as ``classical.'' The results from these different input samples are all consistent (within 68\% confidence) with the original full sample of 38 satellites.}}
    \label{fig:othersamples}
\end{figure}

\begin{figure*}
    \includegraphics[width=\linewidth]{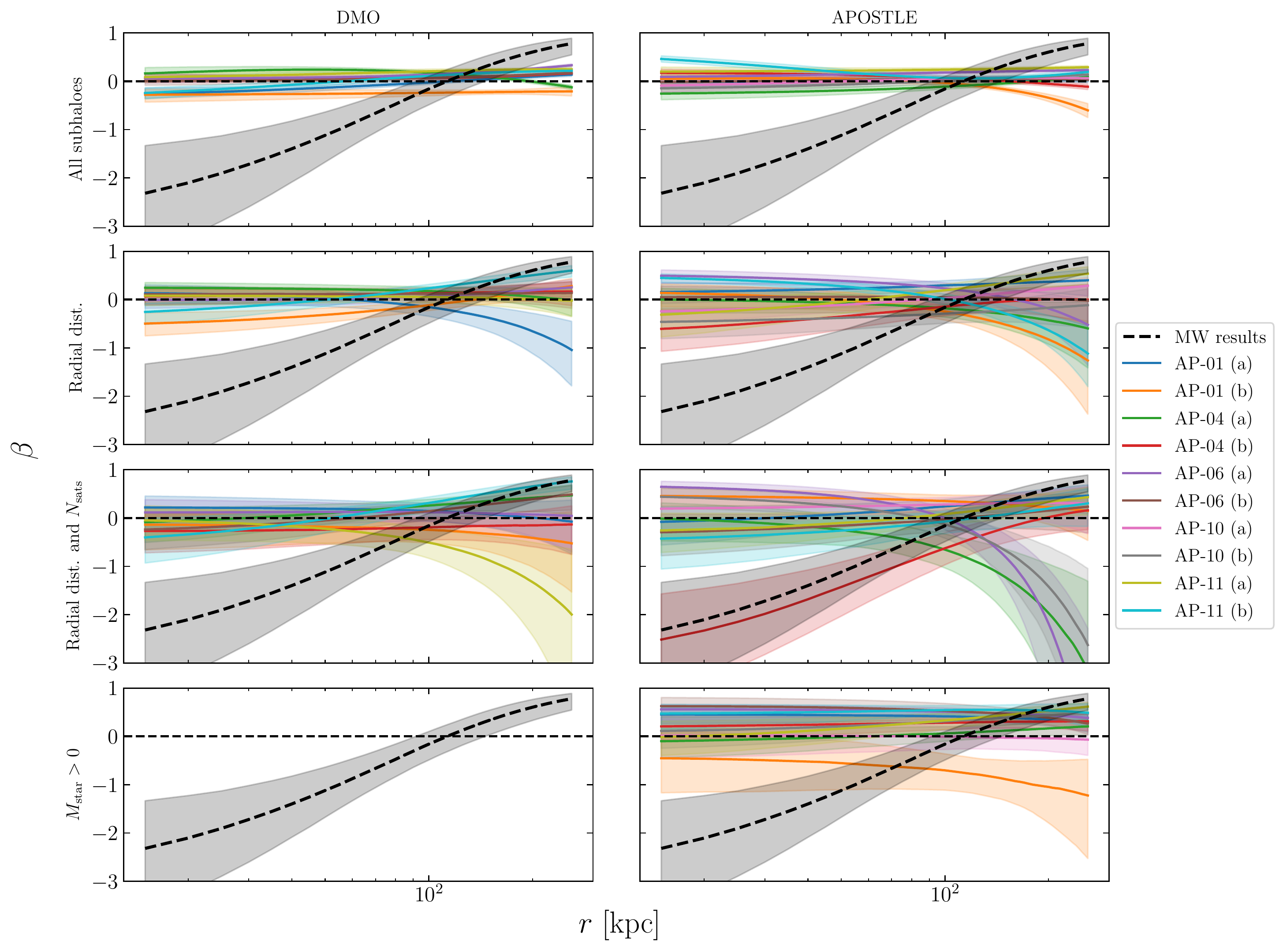}
    \caption{Posterior distribution for the simulated \Ref{APOSTLE} systems assuming the parametrized $\beta(r)$ model. \Ref{The left column corresponds dark matter-only runs of the APOSTLE volumes, while the right column to APOSTLE with baryons.} From top to bottom the rows correspond to different subhalo populations: the full sample of subhaloes ($V_\text{max} > 5$~km~s$^{-1}$), matching the radial distribution of MW satellites, matching both the radial distribution and number (see Section \ref{subsec:matchdata} for details on matching the radial distribution/abundance), and $M_\text{star} > 0$ (radial matching is not applied in the final row, $M_\text{star} > 0$). For comparison, the result for the MW satellite system is shown in \Ref{grey}, with a \Ref{black} dashed curve, in each panel. The different simulated systems are shown with different colors, with the shaded regions corresponding to the $16-84$\% confidence intervals. \Ref{Note that each halo in an APOSTLE pair is treated separately, with the more massive halo denoted as (a) and the less massive one as (b). Additionally, note that AP-10 does not have a completed high-resolution DMO run, so there are two fewer curves in the left panel than the right.} The horizontal black dashed line corresponds to the isotropic case $\beta=0$. \Ref{The DMO and APOSTLE systems generally do not exhibit a dip in the $\beta(r)$ profile, but as the simulations are convolved with the observed MW radial distribution and abundance of satellites (first three rows, top to bottom), some APOSTLE systems can exhibit a dip in their $\beta(r)$ profile.}}
    \label{fig:apostle}
\end{figure*}

\par Using the \Ref{other} proper motion samples described in Section~\ref{subsec:PMstudy} does not impact the results; the ``gold'' sample and ``gold'' satellites with \citetalias{fritz} proper motions have nearly the same $\beta(r)$ profile. Furthermore, this dip in $\beta$ does not appear to be dependent on a particular satellite or population of satellites.  We repeated the analysis removing Sagittarius (which has a well-constrained proper motion at $r\sim18$ kpc), removing satellites with luminosities above or below the median luminosity, and removing candidate satellites of the LMC identified by \citet{kallivayalil}: Horologium I, Carina II, Carina III, and Hydrus I. \Ref{The results from these different input samples are all consistent (within 68\% confidence) with the original full sample of 38 satellites (see Figure \ref{fig:othersamples}).}

\par Taken together, these results indicate that satellites closer to the Galactic centre have more tangentially-biased (near-circular) motions than those farther away. This dip in $\beta(r)$ could be a reflection of the destruction of substructure by the central stellar disc, as discussed by \citet{2017MNRAS.467.4383S} and \citet{2017MNRAS.471.1709G}. To interpret this result for the MW satellite system, we move on to analyse the simulated systems in the APOSTLE and Auriga suites using the same methods.

\begin{figure*}
    \includegraphics[width=\linewidth]{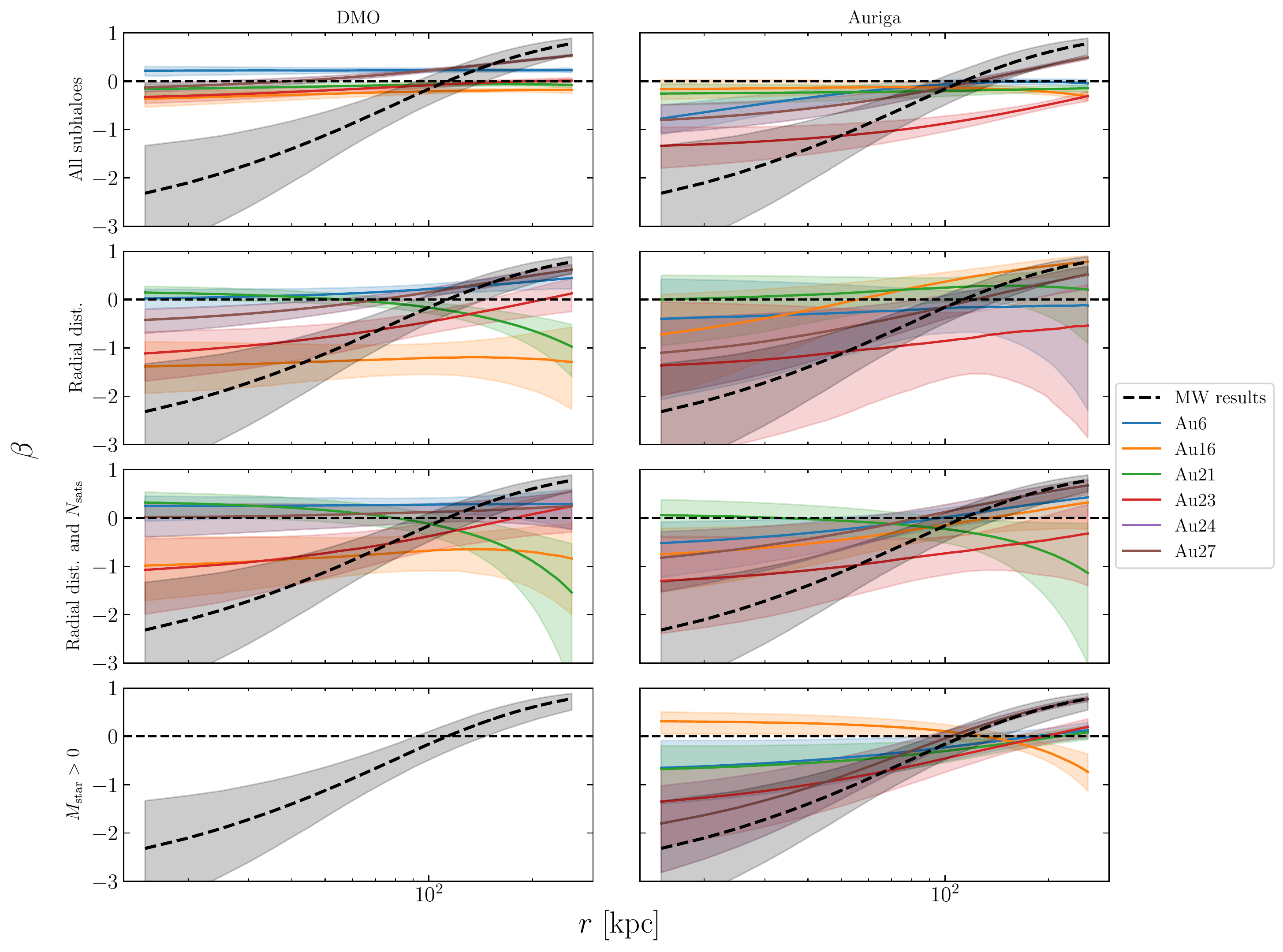}
    \caption{\Ref{Same as Figure \ref{fig:apostle}, but for simulated Auriga systems. The dip in the $\beta(r)$ profile for Auriga hydrodynamic simulations is far more pronounced than in DMO or APOSTLE hydrodynamic simulations, likely due to the formation of a central stellar disc.}}
    \label{fig:auriga}
\end{figure*}

\subsection{Simulations}

\par The posterior distributions for the $\beta(r)$ profiles of simulated MW analogs are shown in \Ref{Figure~\ref{fig:apostle} for APOSTLE and Figure~\ref{fig:auriga} for Auriga}. When considering all subhaloes with $V_\text{max} > 5$~km~s$^{-1}$ (top row), it is clear that the presence of a massive stellar disc affects the radial $\beta$ profile.  The $\beta(r)$ profiles for DMO hosts are nearly flat at $\beta\sim0$, indicating isotropic motions at all radii. The inclusion of baryons in APOSTLE does not have a noticeable effect on the $\beta(r)$ profiles, which are very similar to DMO. Only the Auriga haloes exhibit a dip in the $\beta$ profile near the centre, resulting from the massive central disc preferentially destroying radial orbits that pass near the galaxy.

\par However, the $\beta(r)$ profile estimates of our simulated systems are sensitive to the radial distribution of the subhaloes. Matching the radial distribution of subhaloes with that of the MW satellites, following the procedure described in Section~\ref{subsec:matchdata}, results in similar $\beta(r)$ estimates in some systems that do not contain stellar discs; the estimates for some of the APOSTLE and even DMO systems become consistent with the results for the MW satellites (second row). This similarity is even more pronounced when the subhaloes are also selected to match the total number of MW satellites in addition to the radial distribution (third row).

\par While these corrections bring some DMO and APOSTLE estimates of $\beta(r)$ in line with that of the MW, the Auriga systems still provide the best agreement. There are many corrected DMO and APOSTLE systems that still have $\beta\sim0$ near the centre, but only a few corrected Auriga profiles that do not have a dip in $\beta$. These results suggest that the dip in the $\beta$ profile for the MW satellite system is likely best explained by effects due to the stellar disc, but also is sensitive to the radial distribution of tracers considered.

\par Finally, we also consider the sample of subhaloes in APOSTLE and Auriga that contain stars at $z=0$ ($M_\ast > 0$, bottom row). Nearly every $\beta(r)$ profile matches that of the full subhalo population, albeit with increased scatter due to a smaller sample size (see Table~\ref{tab:simprops} for the number in each population). The only exception is the less massive halo of AP-01, whose $\beta$ profile is shifted to lower values at all radii but maintains the same shape. The agreement between the $\beta(r)$ profiles when considering all subhaloes {\em vs.}~the subsample containing stars suggests that the $\beta$ profile (corrected for observational distance biases) traced by the MW satellites is likely indicative of the intrinsic profile for the MW, unaffected by the complex physics that dictate which subhaloes are populated by satellites \citep{2015MNRAS.448.2941S, 2017MNRAS.464.3108G, 2019ApJ...873...34N}.

\section{Discussion} \label{sec:discussion}

\par It is clear from our results that dwarf galaxy satellites closer to the centre of the Milky Way have tangentially-biased motions while those farther from the centre have radially-biased motions. In this Section we explore some interpretations of the $\beta(r)$ profile and place our results in the context of those from other tracers.

\subsection{Stellar disc}

\par The inclusion of baryonic processes in cosmological simulations has helped resolve a number of small-scale challenges to \LCDM.  A notable effect is the destruction of substructure due to the potential of a massive stellar disc. \citet{2014ApJ...786...87B} found that 6 of the 8 subhaloes in a DMO simulation that did not have a baryonic simulation counterpart had pericentric passages that took them within 30 kpc of the galaxy centre. \citet{2017MNRAS.467.4383S} found that the presence of baryons near the centre of APOSTLE haloes reduces the number of subhaloes by factors of $\sim1/4 - 1/2$ independently of subhalo mass but increasingly towards the host halo centre. \citet{2017MNRAS.471.1709G} found similar destruction of subhaloes in the Latte simulation suite and showed that simply embedding a central disc potential in DMO simulations reproduced these radial subhalo depletion trends, arguing that the additional tidal field from the central galaxy is the primary cause of subhalo depletion \citep[see also][]{2010ApJ...709.1138D, 2015MNRAS.452.2367Y, 2017MNRAS.465L..59E}. We also note that \citet{2016MNRAS.458.1559Z} found similar results in simulations using the AREPO code, the same code with which Auriga was performed.

\par A central stellar disc, whether artificially embedded in DMO simulations or formed through the inclusion of baryonic physics, preferentially destroys subhaloes on radial orbits that pass close to the disc. The surviving population then has tangentially-biased motions compared to DMO \citep{2017MNRAS.467.4383S, 2017MNRAS.471.1709G}, which is expected to be reflected in a lower value of $\beta$. However, there is also a radial dependence of $\beta$ which has not yet been explored; with increasing distance from the central galaxy the destructive effects of the disc potential weaken, causing the $\beta(r)$ profile to rise to $\beta \sim 0.5$ in the outer region ($100\text{ kpc} < r < R_\text{vir}$) as subhaloes are more likely to be on their first infall \citep{2004MNRAS.352..535D, 2010MNRAS.402...21N}. Additionally, as surviving massive satellites pass near the stellar disc, both experience a torque and exchange angular momentum, likely inducing further circularization of the surviving satellite orbits \citep{2017MNRAS.465.3446G, 2017MNRAS.472.3722G}.

\par Our simulation results are consistent with this interpretation. When considering all subhaloes with $V_\text{max} > 5$~km~s$^{-1}$ in the APOSTLE and Auriga suites (top rows of \Ref{Figures~\ref{fig:apostle} and \ref{fig:auriga}}), the $\beta(r)$ profiles for DMO and APOSTLE haloes, which have less massive central galaxies, are relatively constant with $\beta \gtrsim 0$.  In stark contrast, the $\beta(r)$ profiles for Auriga haloes have $\beta \lesssim -0.5$ near the centre of the halo and increase to $\beta \gtrsim 0$ by $\sim 200$ kpc.  These results are similar when considering, instead, subhaloes that contain stars at $z=0$ (bottom rows of \Ref{Figures~\ref{fig:apostle} and \ref{fig:auriga}}). 

\par The Auriga simulations produce stellar discs that are massive, thin, and radially extended, like that of the MW, while APOSTLE forms less massive host galaxies with weaker discs. This distinction impacts the orbital distribution of subhaloes and results in the Auriga subhaloes showing a variation of $\beta$ similar to that of the MW satellite system.

\par \Ref{However, it is worth noting the possibility that not all of the subhalo disruption is due to the physical effects of the stellar disc. \citet{2018MNRAS.474.3043V} and \citet{2018MNRAS.475.4066V} raise concerns about artificial subhalo disruption due to numerical effects, suggesting that tracking subhalo disruption requires many more particles than required for typical simulation convergence tests. For example, \citet{2018MNRAS.474.3043V} find that orbits passing within 10-20\% of the virial radius of a host may require $N > 10^6$ particles for an accurate treatment. More work may be required to understand the differences between these results and those from typical convergence tests. This will, in turn, inform our understanding of how much subhalo disruption is due to physical effects of the stellar disc vs.~numerical effects and how this impacts the inferred $\beta(r)$ profile.}

\subsection{The radial distribution}

\par This clean interpretation of a $\beta(r)$ profile caused by the tidal field of the central galaxy becomes muddier when accounting for the observed radial distribution of the Milky Way satellites. We know the current census of satellites is incomplete both radially, due to surface brightness and luminosity selection effects, and in area on the sky, as less than half of the sky has been covered by surveys capable of finding ultra-faint satellite galaxies \citep{2018PhRvL.121u1302K, 2018MNRAS.479.2853N}. This results in a satellite sample that is more centrally concentrated than those found in M31 and in cosmological simulations (\citealt{2014MNRAS.439...73Y, 2018arXiv180803654G, 2018arXiv181112413K}, however, see \citealt{2019MNRAS.483.2000L}), giving greater weight to satellites located closer to the centre.

\par We attempt to account for this by matching the abundance and/or radial distribution of simulated subhaloes with $V_\text{max} > 5$~km~s$^{-1}$ to that of the MW satellites.  Applying these corrections to simulated MW-mass systems tends to lower $\beta$ estimates relative to when the full population is used (see \Ref{Figures~\ref{fig:apostle} and \ref{fig:auriga}, middle two rows}). As a result, the inferred $\beta(r)$ profiles for some DMO and APOSTLE haloes, which do not contain massive central galaxies, are consistent with that of the MW satellites.

\par This is not to say that the impact of the central disc is not crucial to explaining the anisotropy of the MW satellite system. As shown in \Ref{Figures~\ref{fig:apostle} and \ref{fig:auriga}}, for any given selection criterion applied to the subhaloes the dip in the $\beta(r)$ profiles is most prominent for the Auriga host haloes, which have massive central discs. However, a more complete analysis of the MW disc's impact on the $\beta(r)$ profile would require understanding the true selection function for the MW satellites. Knowing this selection function, combined with a more \Ref{detailed} modeling procedure \Ref{(e.g.~using a distribution function in action/angle coordinates, as in \citealt{2019MNRAS.484.2832V}, would lend greater insight to orbital properties at the expense of assuming a MW potential)}, would enhance future studies of the $\beta(r)$ profile for the MW satellite galaxies.

\subsection{Comparison with other tracers} \label{subsec:othertracers}

\begin{figure}
    \includegraphics[width=\linewidth]{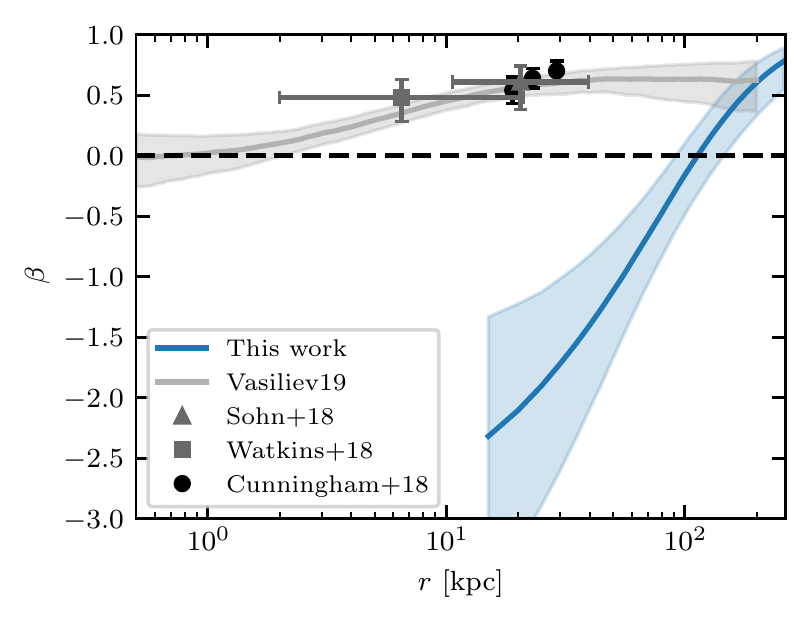}
    \caption{Comparison of MW $\beta(r)$ results between different tracers. The \Ref{grey} points and contours correspond to studies using globular clusters, while the black points correspond to studies using halo stars. The blue contours are the results from this work. The horizontal black dashed line corresponds to the isotropic case, $\beta=0$.}
    \label{fig:tracers}
\end{figure}

\par Finally, it is interesting to compare our $\beta(r)$ results with those using other tracers of the MW potential (see Figure~\ref{fig:tracers}). \citet{2018arXiv181012201C} used HST proper motions of $N\sim200$ halo stars in four different fields, spherically averaging to find $\beta\sim0.5-0.7$ over $19 < r < 29$ kpc. This is higher than the values found in several other studies using line-of-sight velocities alone to constrain the anisotropy of the stellar halo, which tend to prefer isotropic or tangentially-biased $\beta$ values \citep[for a summary, see Figure~6 in][]{2016ApJ...820...18C}.

\par Using MW globular clusters (GCs), \citet{2018ApJ...862...52S} estimated $\beta = 0.609_{-0.229}^{+0.130}$ over $10.6 < r < 39.5$ kpc with proper motions from HST while \citet{2019ApJ...873..118W} found $\beta = 0.48_{-0.20}^{+0.15}$ over $2.0 < r < 21.1$ kpc with proper motions from {\it Gaia} DR2. These two values suggest a trend for the GC orbits to become more radially-biased with increasing distance. Indeed, \citet{2019MNRAS.484.2832V} modelled $\beta(r)$ for the MW GCs using a distribution function-based method in action/angle space and found a steady increase from $\beta\sim0.0$ at 0.5 kpc to $\beta\sim0.6$ at 200 kpc \citep[see Figure~7 of][]{2019MNRAS.484.2832V}, consistent with these other results. 

\par The dip in the $\beta(r)$ profile for the MW globular cluster system detected by \citet{2019MNRAS.484.2832V} is qualitatively similar in shape to what we find for the MW dwarf satellites, but is very different both in characteristic radial scale and in overall amplitude. At $r>100$~kpc the inferred values of $\beta$ are similar. \Ref{The dip in the globular cluster profile may possibly be attributed to the accretion history vs.~in situ formation \citep{2001ApJ...561..751F, 2008ApJ...689..919P}.} It is \Ref{also} possible that both the globular clusters and stellar halo are remnants of stars previously attached to subhaloes on radial orbits, which are preferentially destroyed by the stellar disc, and maintain the anisotropy of their progenitors (see \citealt{2001ApJ...548...33B, 2005ApJ...635..931B, 2008ApJ...680..295B} for halo stars; \citealt{1984ApJ...277..470P, 2006MNRAS.368..563M, 2017MNRAS.472.3120B} for globular clusters). This potential connection between different tracers of the MW $\beta(r)$ profile merits further modeling, possibly with a joint analysis of Milky Way halo stars, globular clusters, and dwarf galaxies.

\section{Summary} \label{sec:conclusions}

\par In this work we have analysed the kinematics of 38 Milky Way satellites focusing on an estimate of the velocity anisotropy parameter, $\beta$, and its dynamical interpretation. Utilizing the latest satellite proper motions inferred from {\it Gaia} DR2 data, we modelled $\beta$ using a likelihood method and, for the first time, estimated $\beta(r)$ for the MW satellite system. We then compared these results with expectations from \LCDM\ using the APOSTLE and Auriga simulation suites. A summary of our main results is as follows:
\begin{itemize}
    \item The MW satellites have overall tangentially-biased motions, with best-fitting \Ref{constant} $\beta = -1.02_{-0.45}^{+0.37}$. By parametrizing $\beta(r)$, we find that the anisotropy profile for the MW satellite system increases from $\beta\sim-2$ at $r\sim20$~kpc to $\beta\sim0.5$ at $r\sim200$~kpc, indicating that satellites closer to the Galactic centre have tangentially-biased motions while those farther out have radially-biased motions.
    \item Comparing these results with the APOSTLE and Auriga galaxy formation simulations, we find that satellites surrounding the massive and radially extended stellar discs formed in Auriga have similar $\beta(r)$ profiles to that of the MW, while the weaker discs in APOSTLE produce profiles that are similar to those from DMO simulations. This suggests that the central stellar disc affects the $\beta(r)$ profile of the MW satellite system by preferentially destroying radial orbits that pass near the disc, as discussed by \citet{2017MNRAS.467.4383S} and \citet{2017MNRAS.471.1709G}.
    \item However, when matching the radial distributions of simulated subhaloes to that of the MW satellites, some of the inferred $\beta(r)$ profiles for APOSTLE and even DMO haloes also can match the MW data. This implies that the partial sky coverage and the increasing incompleteness with distance of the currently available satellite sample significantly impair the ability of our scheme to robustly estimate the true $\beta(r)$ profile.
\end{itemize}

\par The difficulty in interpreting the inferred $\beta(r)$ profile may also be alleviated by more fully exploring the Milky Way's virial volume. \citet{2018MNRAS.479.2853N} expect that -- assuming a MW halo mass of $1.0\times10^{12}$ M$_\odot$ -- there are $46_{-8}^{+12}$ ultra-faint \Post{($-8 < M_V \leq -3$)} satellites and $61_{-23}^{+37}$ hyper-faint ($-3 < M_V \leq 0$) satellites within 300 kpc that are detectable. At least half of these satellites should be found by LSST within the next decade. 

\par Obtaining proper motions for these faint and distant objects will be challenging but clearly possible, given the results already obtained for 7 satellites fainter than $M_V = -5$ and farther than $d_\odot = 100$~kpc.  Furthermore, since the precision in proper motion measurements grows as the 1.5 power of the time baseline, the satellite proper motions from {\it Gaia} should be a factor 4.5 more precise after the nominal mission and possibly a factor 12 more precise after the extended mission \citepalias{fritz}. \Ref{Artificially scaling the observed proper motion errors by a factor of 4.5 results in a $\beta(r)$ profile that has a narrower confidence interval by a factor of $\sim$1.5 and provides better agreement with the Auriga $\beta(r)$ profiles.} With this improved dataset, future studies will be less limited by observational selection effects and be able to study in greater depth the impact of the central stellar disc on the $\beta(r)$ profile of the Milky Way satellite system.

\section*{Acknowledgements}

\par We thank Marius Cautun for useful discussions regarding this work \Ref{and an anonymous referee for helpful suggestions regarding its presentation}, as well as Emily Cunningham and Eugene Vasiliev for sharing their data. This research made use of the Python Programming Language, along with many community-developed or maintained software packages including Astropy \citep{astropy13, astropy18}, corner.py \citep{corner}, emcee \citep{emcee}, Jupyter \citep{jupyter}, Matplotlib \citep{matplotlib}, NumPy \citep{numpy}, Pandas \citep{pandas}, and SciPy \citep{scipy}. This research made extensive use of \href{https://arxiv.org/}{arXiv.org} and NASA's Astrophysics Data System for bibliographic information.

\par AHR acknowledges support from a Texas A\&M University Merit Fellowship. AF is supported by a European Union COFUND/Durham Junior Research Fellowship (under EU grant agreement no. 609412). ABP acknowledges generous support from the Texas A\&M University and the George P. and Cynthia Woods Institute for Fundamental Physics and Astronomy. LES acknowledges support from DOE Grant de-sc0010813. CSF acknowledges a European Research Council (ERC) Advanced Investigator grant (DMIDAS; GA 786910). FAG acknowledges financial support from FONDECYT Regular 1181264, and funding from the Max Planck Society through a Partner Group grant. RG acknowledges support by the Deutsche Forschungsgemeinschaft (DFG) Research Centre Sonderforschungsbereich (SFB)-881 `The Milky Way System' through project A1.

\par \Post{This work partly used the computing and
storage hardware provided by WestGrid (\href{https://www.westgrid.ca/}{www.westgrid.ca}) and Compute Canada Calcul Canada (\href{https://www.computecanada.ca/}{www.computecanada.ca}), as well as UK Research Data Facility (\href{http://www.archer.ac.uk/documentation/rdf-guide/}{http://www.archer.ac.uk/documentation/rdf-guide}).} Part of the simulations of this paper used the SuperMUC system at the Leibniz Computing Centre, Garching, under the project PR85JE of the Gauss Centre for Supercomputing. This work was supported in part by the UK Science and Technology Facilities Council (STFC) ST/P000541/1. This work used the DiRAC Data Centric system at Durham University, operated by the ICC on behalf of the STFC DiRAC HPC Facility (\href{https://dirac.ac.uk/}{www.dirac.ac.uk}). This equipment was funded by BIS National E-infrastructure capital grant ST/K00042X/1, STFC capital grant ST/H008519/1, and STFC DiRAC Operations grant ST/K003267/1 and Durham University. DiRAC is part of the National E-Infrastructure.

\bibliographystyle{mnras}
\bibliography{references,software}


\appendix

\section{MW satellite properties}

\par Table \ref{tab:properties} lists the observed properties of the MW satellites used throughout this analysis.  Table \ref{tab:galactoprops} lists the Galactocentric spherical positions and velocities, along with corresponding uncertainties, of each satellite obtained by the Monte Carlo sampling detailed in Section \ref{subsec:coords}.

\par The references in the last two columns of Table \ref{tab:properties} are as follows: [1] \citet{helmi}; [2] \citet{simon}; [3] \citet{kallivayalil}; [4] \citet{massari}; [5] \citet{pace}; [6] \citet{2016MNRAS.463..712T}; [7] \citet{2006ApJ...653L.109D}; [8] \citet{2008ApJ...684.1075M}; [9] \citet{2011ApJ...736..146K}; [10] \citet{2008ApJ...688..245W}; [11] \citet{2009ApJ...690..453K}; [12] \citet{2008ApJ...674L..81K}; [13] \citet{2007ApJ...670..313S}; [14] \citet{2008ApJ...675L..73G}; [15] \citet{2012ApJ...756...79S}; [16] \citet{2015AJ....150...90K}; [17] \citet{2014MNRAS.444.3139M}; [18] \citet{2009AJ....137.3100W}; [19] \citet{2018MNRAS.475.5085T}; [20] \citet{2018ApJ...857..145L}; [21] \citet{2010AJ....140..138M}; [22] \citet{2009ApJ...695L..83M}; [23] \citet{2017ApJ...839...20C}; [24] \citet{2016MNRAS.459.2370T}; [25] \citet{2015MNRAS.448.2717W}; [26] \citet{2004AJ....127..861B}; [27] \citet{2015ApJ...813...44L}; [28] \citet{2009AJ....138..459P}; [29] \citet{2015ApJ...805..130K}; [30] \citet{2016ApJ...819...53W}; [31] \citet{aden}; [32] \citet{2012ApJ...756..121M}; [33] \citet{2015ApJ...811...62K}; [34] \citet{2015ApJ...810...56K}; [35] \citet{2016AJ....151..118V}; [36] \citet{2015ApJ...804L...5M}; [37] \citet{2018MNRAS.479.5343K}; [38] \citet{2008ApJ...675..201M}; [39] \citet{2014PASP..126..616S}; [40] \citet{2017ApJ...836..202S}; [41] \citet{2005MNRAS.360..185B}; [42] \citet{2009ApJ...699L.125M}; [43] \citet{2017MNRAS.467..573C}; [44] \citet{2017ApJ...845L..10M}; [45] \citet{2012AJ....144....4M}; [46] \citet{2015ApJ...808...95S}; [47] \citet{2018ApJ...863...25M}; [48] \citet{2015MNRAS.454.1509M}; [49] \citet{2007ApJ...654..897B}; [50] \citet{2011ApJ...733...46S}; [51] \citet{2013AJ....146...94B}; [52] \citet{2013ApJ...770...16K}; [53] \citet{2017MNRAS.467..208O}; [54] \citet{2017ApJ...838...83K}; [55] \citet{2017AJ....154..267C}; [56] \citet{2015ApJ...807...50B}; [57] \citet{2017ApJ...838...11S}; [58] \citet{2015ApJ...813..109D}; [59] \citet{2013ApJ...767...62G}; [60] \citet{2012ApJ...752...42D}; [61] \citet{2002AJ....124.3222B}; [62] \citet{2011AJ....142..128W}.

\onecolumn
\begin{landscape}
\begin{table*}
\caption{Properties of the satellites used in this analysis: RA, Dec, absolute magnitude, heliocentric distance, line-of-sight velocity, proper motion from \citet{fritz}, proper motion for the ``gold'' sample described in Section \ref{subsec:PMstudy}, and study used for the ``gold'' PM.}
\centering
	\begin{tabular}{lcrrcccccccr}
	\hline \hline
    Satellite & RA & Dec & $M_V$ & $d_\odot$ & $v_\odot$ & $\mu_{\alpha^\ast}^\text{F18}$ & $\mu_{\delta}^\text{F18}$ & $\mu_{\alpha^\ast}^\text{gold}$ & $\mu_{\delta}^\text{gold}$ & PM$_\text{gold}$ & Refs. \\
    & [deg] & [deg] & [mag] & \Ref{[kpc]} & [km~s$^{-1}$] & [mas yr$^{-1}$] & [mas yr$^{-1}$] & [mas yr$^{-1}$] & [mas yr$^{-1}$] &  &  \\
    \hline
    Aquarius II & 338.481 & -9.327 & -4.36 & $107.9 \pm 3.3$ & $-71.1 \pm 2.5$ & $-0.252 \pm 0.53$ & $0.011 \pm 0.452$ & $-0.491 \pm 0.306$ & $-0.049 \pm 0.266$ & [3] & [6] \\
    Bo{\"o}tes I & 210.015 & 14.512 & -6.3 & $66 \pm 3.0$ & $102.2 \pm 0.8$ & $-0.554 \pm 0.098$ & $-1.111 \pm 0.076$ & $-0.459 \pm 0.041$ & $-1.064 \pm 0.029$ & [1] & [7] [8] [9] \\
    Bo{\"o}tes II & 209.521 & 12.859 & -2.3 & $42 \pm 1.6$ & $-117.1 \pm 7.6$ & $-2.686 \pm 0.393$ & $-0.53 \pm 0.292$ & $-2.517 \pm 0.325$ & $-0.602 \pm 0.235$ & [2] & [10] [11] \\
    Canes Venatici I & 202.016 & 33.559 & -8.6 & $210 \pm 6$ & $30.9 \pm 0.6$ & $-0.159 \pm 0.1$ & $-0.067 \pm 0.064$ & \ldots & \ldots & \ldots & [8] [12] [13] \\
    Canes Venatici II & 194.292 & 34.321 & -4.6 & $160 \pm 7$ & $-128.9 \pm 1.2$ & $-0.342 \pm 0.238$ & $-0.473 \pm 0.178$ & \ldots & \ldots & \ldots & [13] [14] [15] \\
    Carina I & 100.407 & -50.966 & -8.6 & $105.6 \pm 5.4$ & $222.9 \pm 0.1$ & $0.485 \pm 0.038$ & $0.131 \pm 0.038$ & $0.495 \pm 0.015$ & $0.143 \pm 0.014$ & [1] & [16] [17] [18] \\
    Carina II & 114.107 & -57.999 & -4.5 & $37.4 \pm 0.4$ & $477.2 \pm 1.2$ & $1.867 \pm 0.085$ & $0.082 \pm 0.08$ & $1.79 \pm 0.06$ & $0.01 \pm 0.05$ & [4] & [19] [20] \\
    Carina III & 114.630 & -57.900 & -2.4 & $27.8 \pm 0.6$ & $284.6 \pm 3.25$ & $3.046 \pm 0.132$ & $1.565 \pm 0.147$ & $3.065 \pm 0.095$ & $1.567 \pm 0.104$ & [3] & [19] [20] \\
    Coma Berenices I & 186.746 & 23.908 & -3.8 & $42 \pm 1.5$ & $98.1 \pm 0.9$ & $0.471 \pm 0.113$ & $-1.716 \pm 0.11$ & $0.546 \pm 0.092$ & $-1.726 \pm 0.086$ & [2] & [13] [21] [22] \\
    Crater II & 177.310 & -18.413 & -8.2 & $117.5 \pm 1.1$ & $87.5 \pm 0.4$ & $-0.184 \pm 0.07$ & $-0.106 \pm 0.068$ & $-0.246 \pm 0.052$ & $-0.227 \pm 0.026$ & [3] & [23] [24] \\
    Draco I & 260.060 & 57.965 & -8.75 & $76 \pm 6$ & $-291.0 \pm 0.1$ & $-0.012 \pm 0.013$ & $-0.158 \pm 0.038$ & $-0.019 \pm 0.009$ & $-0.145 \pm 0.01$ & [1] & [25] [26] \\
    Draco II & 238.198 & 64.565 & -2.9 & $20 \pm 3.0$ & $-347.6 \pm 1.75$ & $1.242 \pm 0.282$ & $0.845 \pm 0.291$ & $1.165 \pm 0.26$ & $0.866 \pm 0.27$ & [3] & [24] [27] \\
    Fornax & 39.962 & -34.511 & -13.4 & $147 \pm 9$ & $55.3 \pm 0.1$ & $0.374 \pm 0.035$ & $-0.401 \pm 0.035$ & $0.376 \pm 0.003$ & $-0.413 \pm 0.003$ & [1] & [18] [28] \\
    Grus I & 344.176 & -50.163 & -3.4 & $120.2 \pm 11.1$ & $-140.5 \pm 2.0$ & $-0.261 \pm 0.178$ & $-0.437 \pm 0.242$ & $-0.25 \pm 0.16$ & $-0.47 \pm 0.23$ & [5] & [29] [30] \\
    Hercules & 247.763 & 12.787 & -6.6 & $132 \pm 6$ & $45.2 \pm 1.09$ & $-0.297 \pm 0.123$ & $-0.329 \pm 0.1$ & \ldots & \ldots & \ldots & [31] [32] \\
    Horologium I & 43.882 & -54.119 & -3.5 & $79 \pm 7$ & $112.8 \pm 2.55$ & $0.891 \pm 0.105$ & $-0.55 \pm 0.099$ & $0.95 \pm 0.07$ & $-0.55 \pm 0.06$ & [5] & [33] \\
    Hydra II & 185.425 & -31.985 & -4.8 & $151 \pm 8$ & $303.1 \pm 1.4$ & $-0.416 \pm 0.523$ & $0.134 \pm 0.426$ & $-0.417 \pm 0.402$ & $0.179 \pm 0.339$ & [3] & [34] [35] [36] \\
    Hydrus I & 37.389 & -79.309 & -4.71 & $27.6 \pm 0.5$ & $80.4 \pm 0.6$ & $3.733 \pm 0.052$ & $-1.605 \pm 0.05$ & $3.761 \pm 0.029$ & $-1.371 \pm 0.027$ & [3] & [37] \\
    Leo I & 152.122 & 12.313 & -12.03 & $258.2 \pm 9.5$ & $282.5 \pm 0.1$ & $-0.086 \pm 0.069$ & $-0.128 \pm 0.071$ & $-0.097 \pm 0.056$ & $-0.091 \pm 0.047$ & [1] & [38] [39] \\
    Leo II & 168.370 & 22.152 & -9.6 & $233 \pm 15$ & $78.5 \pm 0.6$ & $-0.025 \pm 0.087$ & $-0.173 \pm 0.09$ & $-0.064 \pm 0.057$ & $-0.21 \pm 0.054$ & [1] & [40] [41] \\
    Leo IV & 173.233 & -0.540 & -4.97 & $154 \pm 5$ & $132.3 \pm 1.4$ & $-0.59 \pm 0.534$ & $-0.449 \pm 0.362$ & \ldots & \ldots & \ldots & [13] [42] \\
    Leo V & 172.784 & 2.222 & -4.4 & $173 \pm 5$ & $172.1 \pm 2.2$ & $-0.097 \pm 0.56$ & $-0.628 \pm 0.307$ & \ldots & \ldots & \ldots & [43] [44] \\
    LMC & 80.894 & -69.756 & -18.1 & $51.0 \pm 2.0$ & $262.2 \pm 3.4$ & \ldots & \ldots & $1.85 \pm 0.03$ & $0.24 \pm 0.03$ & [1] & [45] \\
    Pisces II & 344.634 & 5.955 & -4.1 & $183 \pm 15$ & $-226.5 \pm 2.7$ & $-0.108 \pm 0.647$ & $-0.586 \pm 0.502$ & \ldots & \ldots & \ldots & [15] [34] \\
    Reticulum II & 53.949 & -54.047 & -3.6 & $31.4 \pm 1.4$ & $64.8 \pm 0.5$ & $2.398 \pm 0.053$ & $-1.319 \pm 0.059$ & $2.36 \pm 0.05$ & $-1.32 \pm 0.06$ & [5] & [46] [47] \\
    Sagittarius I & 283.831 & -30.545 & -13.5 & $26 \pm 2.0$ & $140.0 \pm 2.0$ & $-2.736 \pm 0.036$ & $-1.357 \pm 0.036$ & $-2.692 \pm 0.001$ & $-1.359 \pm 0.001$ & [1] & [45] \\
    Sculptor & 15.039 & -33.709 & -10.7 & $83.9 \pm 1.5$ & $111.4 \pm 0.1$ & $0.084 \pm 0.036$ & $-0.133 \pm 0.0356$ & $0.082 \pm 0.005$ & $-0.131 \pm 0.004$ & [1] & [18] [48] \\
    Segue 1 & 151.763 & 16.074 & -1.5 & $23 \pm 2$ & $208.5 \pm 0.9$ & $-1.697 \pm 0.198$ & $-3.501 \pm 0.178$ & $-1.867 \pm 0.11$ & $-3.282 \pm 0.102$ & [2] & [49] [50] \\
    Segue 2 & 34.817 & 20.175 & -2.5 & $36.6 \pm 2.45$ & $-40.2 \pm 0.9$ & $1.656 \pm 0.161$ & $0.135 \pm 0.113$ & $1.01 \pm 0.14$ & $-0.48 \pm 0.18$ & [4] & [51] [52] \\
    Sextans & 153.268 & -1.620 & -9.3 & $92.5 \pm 2.2$ & $224.2 \pm 0.1$ & $-0.438 \pm 0.045$ & $0.055 \pm 0.045$ & $-0.496 \pm 0.025$ & $0.077 \pm 0.02$ & [1] & [18] [53] \\
    SMC & 13.187 & -72.829 & -16.8 & $64.0 \pm 4.0$ & $145.6 \pm 0.6$ & \ldots & \ldots & $0.797 \pm 0.03$ & $-1.22 \pm 0.03$ & [1] & [45] \\
    Triangulum II & 33.322 & 36.172 & -1.2 & $28.4 \pm 1.6$ & $-381.7 \pm 1.1$ & $0.588 \pm 0.194$ & $0.554 \pm 0.169$ & $0.588 \pm 0.187$ & $0.554 \pm 0.161$ & [3] & [54] [55] \\
    Tucana II & 343.060 & -58.570 & -3.9 & $57.5 \pm 5.3$ & $-129.1 \pm 3.5$ & $0.91 \pm 0.069$ & $-1.159 \pm 0.082$ & $0.91 \pm 0.06$ & $-1.16 \pm 0.08$ & [5] & [30] [56] \\
    Tucana III & 359.150 & -59.600 & -2.4 & $25 \pm 2$ & $-102.3 \pm 0.4$ & $-0.025 \pm 0.049$ & $-1.661 \pm 0.049$ & $-0.03 \pm 0.04$ & $-1.65 \pm 0.04$ & [5] & [57] [58] \\
    Ursa Major I & 158.685 & 51.926 & -6.75 & $97.3 \pm 5.85$ & $-55.3 \pm 1.4$ & $-0.683 \pm 0.1$ & $-0.72 \pm 0.135$ & $-0.659 \pm 0.093$ & $-0.635 \pm 0.131$ & [2] & [13] [59] \\
    Ursa Major II & 132.874 & 63.133 & -3.9 & $34.7 \pm 2.1$ & $-116.5 \pm 1.9$ & $1.691 \pm 0.064$ & $-1.902 \pm 0.075$ & $1.661 \pm 0.053$ & $-1.87 \pm 0.065$ & [2] & [13] [60] \\
    Ursa Minor & 227.242 & 67.222 & -8.4 & $76 \pm 4$ & $-246.9 \pm 0.1$ & $-0.184 \pm 0.044$ & $0.082 \pm 0.042$ & $-0.182 \pm 0.01$ & $0.074 \pm 0.008$ & [1] & [61] \\
    Willman 1 & 162.341 & 51.053 & -2.7 & $38 \pm 7$ & $-12.8 \pm 1.0$ & $0.199 \pm 0.194$ & $-1.342 \pm 0.37$ & $0.382 \pm 0.119$ & $-1.152 \pm 0.216$ & [2] & [8] [62] \\
    \hline \hline
  	\end{tabular}
\label{tab:properties}
\end{table*}
\end{landscape}

\begin{table*}
\centering
    \caption{Galactocentric positions and velocities for each coordinate using \citet{fritz} proper motions \citep[LMC and SMC proper motions from][]{helmi}. Quoted values are the 16\ts{th}, 50\ts{th}, and 84\ts{th} percentiles.}
	\begin{tabular}{lrrrrrrr}
	\hline \hline
    Satellite & $r$ & $\theta$ & $\phi$ & $v_r$ & $v_\theta$ & $v_\phi$ \\
    & [kpc] & [deg] & [deg] & [km~s$^{-1}$] & [km~s$^{-1}$] & [km~s$^{-1}$] \\
    \hline
    Aquarius II & $105.2_{-3.2}^{+3.3}$ & $145.0_{-0.1}^{+0.1}$ & $61.7_{-0.2}^{+0.2}$ & $43.2_{-20.2}^{+19.5}$ & $-282.3_{-239.9}^{+222.8}$ & $29.0_{-252.0}^{+268.5}$ \\
    Bo{\"o}tes I & $63.6_{-2.8}^{+2.9}$ & $13.6_{-0.3}^{+0.3}$ & $-2.9_{-0.1}^{+0.1}$ & $97.5_{-3.4}^{+3.3}$ & $110.7_{-25.4}^{+26.2}$ & $-126.0_{-33.9}^{+32.4}$ \\
    Bo{\"o}tes II & $39.8_{-1.5}^{+1.6}$ & $10.4_{-0.5}^{+0.4}$ & $-13.3_{-0.7}^{+0.6}$ & $-48.1_{-16.2}^{+16.5}$ & $-311.1_{-68.3}^{+70.7}$ & $-246.9_{-67.4}^{+70.5}$ \\
    Canes Venatici I & $209.8_{-5.8}^{+5.9}$ & $9.8_{-0.0}^{+0.0}$ & $86.0_{-0.4}^{+0.5}$ & $83.5_{-3.4}^{+3.4}$ & $93.1_{-85.8}^{+80.0}$ & $83.8_{-86.1}^{+79.6}$ \\
    Canes Venatici II & $160.9_{-7.0}^{+6.7}$ & $8.8_{-0.1}^{+0.1}$ & $130.4_{-0.6}^{+0.7}$ & $-93.2_{-8.9}^{+8.4}$ & $-144.2_{-137.4}^{+127.7}$ & $135.0_{-181.1}^{+171.6}$ \\
    Carina I & $106.9_{-5.2}^{+5.8}$ & $111.8_{-0.0}^{+0.0}$ & $-104.6_{-0.2}^{+0.3}$ & $-4.6_{-2.9}^{+3.0}$ & $-165.0_{-24.4}^{+22.0}$ & $-22.5_{-18.2}^{+17.4}$ \\
    Carina II & $38.3_{-0.4}^{+0.3}$ & $106.7_{-0.0}^{+0.0}$ & $-103.0_{-0.2}^{+0.2}$ & $204.4_{-4.3}^{+4.3}$ & $-228.6_{-15.6}^{+15.5}$ & $195.6_{-13.6}^{+14.3}$ \\
    Carina III & $29.0_{-0.6}^{+0.6}$ & $106.1_{-0.0}^{+0.0}$ & $-107.2_{-0.4}^{+0.4}$ & $46.6_{-6.6}^{+6.4}$ & $-383.6_{-20.1}^{+18.9}$ & $36.1_{-18.1}^{+16.9}$ \\
    Coma Berenices I & $43.2_{-1.4}^{+1.5}$ & $14.9_{-0.4}^{+0.4}$ & $-158.2_{-0.6}^{+0.6}$ & $28.9_{-5.1}^{+4.7}$ & $-252.6_{-24.0}^{+21.9}$ & $104.5_{-25.3}^{+25.6}$ \\
    Crater II & $116.4_{-1.1}^{+1.1}$ & $47.5_{-0.0}^{+0.0}$ & $-82.3_{-0.1}^{+0.1}$ & $-83.7_{-3.6}^{+3.4}$ & $-77.0_{-37.3}^{+38.5}$ & $-24.2_{-39.0}^{+40.5}$ \\
    Draco I & $75.9_{-5.6}^{+6.0}$ & $55.3_{-0.0}^{+0.0}$ & $93.8_{-0.6}^{+0.6}$ & $-88.5_{-2.9}^{+2.8}$ & $124.1_{-5.7}^{+5.7}$ & $-50.7_{-13.5}^{+14.2}$ \\
    Draco II & $22.4_{-2.7}^{+2.8}$ & $52.5_{-1.0}^{+1.2}$ & $125.4_{-3.0}^{+3.5}$ & $-154.4_{-12.1}^{+12.4}$ & $300.5_{-23.8}^{+26.0}$ & $-68.8_{-32.9}^{+31.6}$ \\
    Fornax & $149.5_{-9.0}^{+8.6}$ & $153.9_{-0.1}^{+0.1}$ & $-129.1_{-0.4}^{+0.3}$ & $-40.9_{-1.5}^{+1.5}$ & $-104.5_{-32.9}^{+30.8}$ & $112.5_{-28.3}^{+28.4}$ \\
    Grus I & $116.3_{-10.6}^{+11.5}$ & $151.6_{-0.3}^{+0.3}$ & $-24.5_{-0.4}^{+0.3}$ & $-203.2_{-7.0}^{+7.0}$ & $-187.5_{-135.2}^{+133.9}$ & $123.7_{-113.6}^{+110.8}$ \\
    Hercules & $126.3_{-6.0}^{+6.1}$ & $51.3_{-0.1}^{+0.1}$ & $30.9_{-0.1}^{+0.1}$ & $150.5_{-3.2}^{+3.3}$ & $-10.0_{-70.7}^{+73.4}$ & $-54.8_{-68.2}^{+73.4}$ \\
    Horologium I & $79.4_{-7.0}^{+7.0}$ & $144.4_{-0.1}^{+0.1}$ & $-99.1_{-1.0}^{+0.9}$ & $-33.7_{-5.5}^{+5.2}$ & $-193.4_{-49.0}^{+46.6}$ & $0.6_{-40.1}^{+42.0}$ \\
    Hydra II & $148.3_{-8.3}^{+7.9}$ & $58.9_{-0.0}^{+0.0}$ & $-67.6_{-0.2}^{+0.2}$ & $129.3_{-21.2}^{+21.1}$ & $-164.7_{-287.8}^{+281.8}$ & $-187.8_{-396.2}^{+392.9}$ \\
    Hydrus I & $25.7_{-0.5}^{+0.5}$ & $129.9_{-0.0}^{+0.0}$ & $-84.4_{-0.5}^{+0.5}$ & $-57.2_{-3.3}^{+3.2}$ & $-328.9_{-9.9}^{+9.6}$ & $-161.9_{-8.6}^{+9.0}$ \\
    LMC & $50.3_{-1.9}^{+2.0}$ & $123.3_{-0.0}^{+0.0}$ & $-90.7_{-0.5}^{+0.4}$ & $63.1_{-4.3}^{+4.3}$ & $-310.3_{-18.0}^{+18.4}$ & $-40.9_{-9.3}^{+8.3}$ \\
    Leo I & $261.9_{-9.3}^{+9.2}$ & $41.7_{-0.0}^{+0.0}$ & $-135.8_{-0.1}^{+0.1}$ & $168.6_{-3.1}^{+3.1}$ & $24.4_{-74.3}^{+66.4}$ & $-71.4_{-97.0}^{+101.2}$ \\
    Leo II & $235.2_{-14.5}^{+15.2}$ & $24.1_{-0.1}^{+0.1}$ & $-142.7_{-0.2}^{+0.2}$ & $18.5_{-4.0}^{+3.8}$ & $-72.0_{-88.7}^{+86.1}$ & $-14.0_{-101.7}^{+112.0}$ \\
    Leo IV & $154.7_{-4.9}^{+5.1}$ & $33.8_{-0.0}^{+0.0}$ & $-99.8_{-0.2}^{+0.2}$ & $13.8_{-21.5}^{+20.8}$ & $321.4_{-270.9}^{+265.2}$ & $-183.6_{-393.5}^{+372.5}$ \\
    Leo V & $174.0_{-5.0}^{+4.6}$ & $31.9_{-0.0}^{+0.0}$ & $-102.9_{-0.2}^{+0.2}$ & $40.5_{-18.9}^{+19.9}$ & $225.1_{-351.4}^{+373.5}$ & $236.3_{-381.6}^{+369.4}$ \\
    Pisces II & $181.8_{-14.5}^{+14.6}$ & $137.4_{-0.0}^{+0.0}$ & $83.1_{-0.3}^{+0.3}$ & $-79.7_{-24.1}^{+24.4}$ & $173.6_{-471.3}^{+475.1}$ & $-356.8_{-533.7}^{+537.6}$ \\
    Reticulum II & $32.8_{-1.3}^{+1.4}$ & $136.9_{-0.2}^{+0.2}$ & $-115.3_{-0.9}^{+0.8}$ & $-99.8_{-3.1}^{+3.0}$ & $-215.8_{-19.1}^{+18.7}$ & $56.4_{-9.8}^{+10.7}$ \\
    SMC & $61.3_{-3.8}^{+4.2}$ & $136.9_{-0.1}^{+0.1}$ & $-66.8_{-0.7}^{+0.6}$ & $-5.6_{-2.4}^{+2.3}$ & $-245.3_{-27.0}^{+26.3}$ & $-67.5_{-17.1}^{+16.2}$ \\
    Sagittarius I & $18.3_{-2.0}^{+2.0}$ & $110.6_{-0.6}^{+0.8}$ & $8.2_{-0.3}^{+0.3}$ & $140.0_{-2.6}^{+2.3}$ & $-275.2_{-17.0}^{+17.2}$ & $-53.3_{-21.4}^{+21.6}$ \\
    Sculptor & $84.0_{-1.5}^{+1.5}$ & $172.5_{-0.1}^{+0.1}$ & $-119.7_{-0.8}^{+0.9}$ & $75.0_{-1.6}^{+1.6}$ & $169.6_{-14.4}^{+13.9}$ & $-72.8_{-15.1}^{+16.2}$ \\
    Segue 1 & $27.9_{-1.9}^{+1.9}$ & $50.4_{-0.8}^{+0.9}$ & $-153.8_{-0.9}^{+0.8}$ & $116.8_{-5.7}^{+5.9}$ & $142.1_{-31.3}^{+34.5}$ & $142.0_{-30.2}^{+34.0}$ \\
    Segue 2 & $42.4_{-2.3}^{+2.4}$ & $121.9_{-0.4}^{+0.3}$ & $156.1_{-0.3}^{+0.4}$ & $72.8_{-4.5}^{+4.6}$ & $-214.7_{-25.2}^{+26.2}$ & $9.5_{-30.1}^{+28.7}$ \\
    Sextans & $95.5_{-2.2}^{+2.3}$ & $49.3_{-0.0}^{+0.0}$ & $-122.2_{-0.1}^{+0.1}$ & $79.2_{-2.6}^{+2.6}$ & $-12.2_{-18.3}^{+17.6}$ & $-239.5_{-21.6}^{+23.1}$ \\
    Triangulum II & $34.7_{-1.6}^{+1.6}$ & $109.2_{-0.3}^{+0.2}$ & $150.0_{-0.4}^{+0.5}$ & $-255.2_{-5.0}^{+4.9}$ & $-175.7_{-24.0}^{+23.6}$ & $-122.6_{-25.5}^{+24.9}$ \\
    Tucana II & $54.0_{-5.2}^{+5.2}$ & $148.1_{-0.5}^{+0.6}$ & $-40.9_{-1.1}^{+0.9}$ & $-187.6_{-4.1}^{+4.6}$ & $48.6_{-18.1}^{+18.3}$ & $-208.0_{-43.1}^{+40.5}$ \\
    Tucana III & $23.0_{-1.9}^{+1.9}$ & $154.5_{-0.3}^{+0.2}$ & $-80.4_{-4.2}^{+3.4}$ & $-228.1_{-2.2}^{+2.3}$ & $28.2_{-21.2}^{+19.3}$ & $48.3_{-13.6}^{+11.0}$ \\
    Ursa Major I & $102.1_{-5.9}^{+5.7}$ & $39.0_{-0.2}^{+0.2}$ & $161.9_{-0.1}^{+0.1}$ & $11.5_{-3.8}^{+3.7}$ & $165.7_{-54.7}^{+54.0}$ & $206.1_{-60.8}^{+62.6}$ \\
    Ursa Major II & $40.9_{-2.0}^{+2.2}$ & $58.8_{-0.3}^{+0.3}$ & $158.6_{-0.3}^{+0.3}$ & $-57.7_{-2.8}^{+2.8}$ & $-280.3_{-24.1}^{+23.7}$ & $32.6_{-17.0}^{+19.9}$ \\
    Ursa Minor & $78.2_{-4.0}^{+4.0}$ & $46.5_{-0.1}^{+0.1}$ & $112.9_{-0.4}^{+0.4}$ & $-71.4_{-2.8}^{+2.7}$ & $136.9_{-12.3}^{+12.8}$ & $-11.5_{-18.4}^{+18.3}$ \\
    Willman 1 & $42.5_{-6.5}^{+6.8}$ & $41.9_{-1.3}^{+1.6}$ & $164.5_{-0.7}^{+0.9}$ & $17.8_{-6.4}^{+6.3}$ & $-106.4_{-59.3}^{+47.6}$ & $-55.5_{-59.6}^{+75.0}$ \\
    \hline \hline
  	\end{tabular}
\label{tab:galactoprops}
\end{table*}

\bsp	
\label{lastpage}
\end{document}